\documentclass[aps, twocolumn, longbibliography, superscriptaddress]{revtex4-2}

\usepackage{amsmath,amssymb,bm,graphicx}
\usepackage{dcolumn}
\usepackage{color}
\usepackage[utf8]{inputenc}
\usepackage[T1]{fontenc}
\usepackage[colorlinks=true,citecolor=blue]{hyperref}
\hypersetup{colorlinks=true,citecolor=blue,linkcolor=blue,urlcolor=blue}
\usepackage{mathtools}

\begin{document}

\title{Polarons in two-dimensional polar materials: All-coupling variational theory}

\author{A.~Kudlis}
\email{andrewkudlis@gmail.com}
\affiliation{Science Institute, University of Iceland, Dunhagi 3, IS-107, Reykjavik, Iceland}
\author{V.~Shahnazaryan}
\email{vanikshahnazaryan@gmail.com}
\affiliation{Abrikosov Center for Theoretical Physics, MIPT, Dolgoprudnyi, Moscow Region 141701, Russia}
\author{I.~V.~Tokatly}
\email{ilya.tokatly@ehu.es}
\affiliation{Departamento de Polímeros y Materiales Avanzados: Física, Química y Tecnología, Universidad del País Vasco, Avenida Tolosa 72, E-20018 San Sebastián, Spain}
\affiliation{Donostia International Physics Center (DIPC), E-20018 Donostia-San Sebastián, Spain}
\affiliation{IKERBASQUE, Basque Foundation for Science, Plaza Euskadi 5, 48009 Bilbao, Spain}

\date{\today}

\begin{abstract}
We present a detailed and self-contained theoretical study of polarons in two-dimensional (2D) polar materials, which extends the classical macroscopic theory of Fr\"ohlich polarons to the 2D case. The theory is fully determined by experimentally accessible parameters, the static and optical 2D polarizabilities of a monolayer, the frequency of transverse optical phonons, and the effective mass of charge carriers. We define a single dimensionless parameter, which characterizes the coupling of electrons with longitudinal optical phonons, analyze both weak- and strong-coupling regimes, and adopt the Feynman variational path-integral approach for a high-quality interpolation between these limits. Our results provide insight into the ground-state energy and effective mass of polarons in the new generation of 2D polar monolayers.
\end{abstract}

\maketitle
%%%%%%%%%%%%%%%%%%%%%%%%%%%%%%%%%%%%%%%%%%%%%%%%%%%%%%%%%%%%%%%%%%%%%%
\allowdisplaybreaks
\section{Introduction}
An electron moving in a polar solid interacts with ions in the lattice sites and locally polarizes the crystal. This leads to the formation of polaron, a quasiparticle composed of an electron surrounded by the local lattice deformation, as schematically shown on Fig.~1. Polarons play an essential role in elucidating the transport and optical properties of polar materials \cite{Devreese-book-1972,alexandrov2010advances,devreese2016fr}. Originally, the idea of polaron was proposed by Landau and further elaborated by Pekar \cite{landau1933electron,pekar1946local,landau1948effective}. In the Landau-Pekar (LP) picture, the polar crystal is viewed as a polarizable medium that adiabatically follows the electron, forming a self-consistent polarization cloud around it. This reduces the electron energy and can lead to its self-trapping in a polarization well. In the 1950s, a complementary picture of an electron dressed by virtual longitudinal optical (LO) phonons was developed by Fr\"ohlich based on a perturbative treatment of the problem \cite{frohlich1950xx}, and somewhat later it was restated in a variational form by Lee, Low, and Pines (LLP) \cite{lee1953motion}. Shortly, Feynman made further significant progress in the theory of polarons by adopting for this purpose a path integral variational formalism \cite{feynman1955slow}. The adiabatic LP and the perturbative Fr\"ohlich approaches describe, respectively, the regimes of strong and weak coupling of an electron to the lattice polarization. In contrast, the Feynman theory is valid for arbitrary interaction strength, providing a high-quality interpolation between the weak and strong coupling limits. Despite its conceptual simplicity, the Feynman approach provides a remarkably accurate description of polarons \cite{Rosenfelder2001}, as has been confirmed by subsequent Monte-Carlo calculations \cite{prokofev1998polaron,mishchenko2000diagrammatic,hahn2018diagrammatic}. In fact, the Feynman theory of polarons is one of the most successful practical applications of the path integral technique.
\begin{figure}
    \centering
    \includegraphics[width=0.8\linewidth]{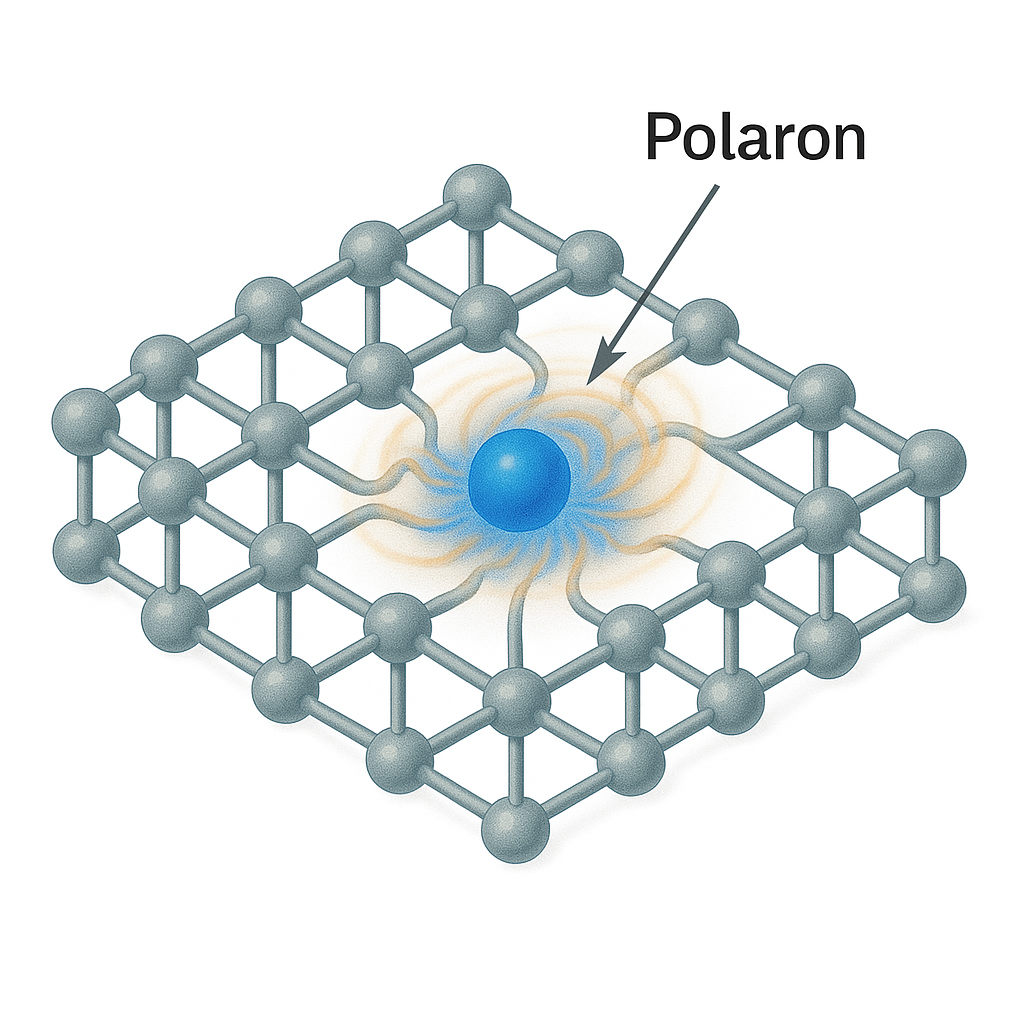}
    \caption{Schematic illustration of a polaron in a polar lattice. The blue sphere represents an electron, whose interaction with the surrounding ions locally deforms the lattice (indicated by the orange halo), effectively dressing the electron with a phonon cloud.}
    \label{fig:enter-label}
\end{figure}

Due to their importance for physics of polar crystals, such as alkali halides and certain oxides, polaron effects have been extensively studied both theoretically and experimentally \cite{devreese2009frohlich,franchini2021polarons}, providing a prototype for strong-coupling electron-phonon physics. Over the past decades, various theoretical techniques, including diagrammatic Monte-Carlo  \cite{prokofev1998polaron,mishchenko2000diagrammatic,hahn2018diagrammatic} and first-principle DFT-based \cite{giustino2017electron,sio2019polarons,sio2019ab} methods, have been used to capture different aspects of polaron physics. An excellent account of the theory of polarons, including an extensive list of literature, can be found in the lecture notes of Ref.~\cite{devreese2016fr}.

Starting from the earlier classical works by Landau, Pekar, Fr\"olich, and Feynman, the formation of polarons has been mostly studied for three-dimensional (3D) polar crystals, first in the bulk and, starting from the 1970s, on the surface \cite{Sak1972,xiaoguang1985exact,peeters1986ground,hahn2018diagrammatic,li2018optical}. Until very recently, the quasi-two-dimensional (2D) polarons, appearing due to the interaction with surface LO phonons on boundaries of 3D crystals, represented the only well-studied example of low-dimensional polarons. The studies of genuine 2D polarons remained relatively sparse, despite rapid progress in the discovery and synthesis of two-dimensional (2D) materials \cite{novoselov20162d}. Experimental reports on 2D polarons exist, for example, in hexagonal boron nitride (hBN) \cite{chen2018emergence}, MoS$_2$ monolayers \cite{kang2018holstein}, and SrTiO$_3$ surfaces \cite{chen2015observation,wang2016tailoring}.  

A key feature of 2D polar materials, relevant for polaron physics, is their reduced, nonlocal dielectric screening \cite{Keldysh1979,Rytova1967,cudazzo2011dielectric}. This very special screening is responsible for non-trivial dispersion of long-wavelength LO phonons and fundamental modifications of electron-phonon interactions \cite{sohier2016two,sohier2017breakdown,sio2022unified,ponce2023long}. Some aspects of polaron physics in 2D polar monolayers have been studied within a first-principles extension of Landau-Pekar theory \cite{sio2023polarons}, and by perturbatively computing polaron shifts of Landau levels \cite{chen2018magneto}. In a recent work \cite{shahnazaryan2025polarons}, we have proposed a macroscopic approach to optical phonon modes and electron-LO phonon coupling in 2D polar materials, which generalizes the classical textbook theory of 3D polar crystals \cite{BornHuang1954,kittel1987quantum}. Similarly to the 3D case, in this approach, the effective electron-phonon Hamiltonian is parametrized solely in terms of macroscopic experimentally accessible quantities –
2D polarizabilities of the monolayer at low and high frequencies. In Ref.~\cite{shahnazaryan2025polarons}, the proposed theory of 2D polar material has been applied to study polarons and polaron-excitons within a variational scheme of LLP, which, strictly speaking, should be valid in the weak coupling regime. 

In the present work, we focus on Fr\"ohlich polarons in 2D polar crystals, building upon a macroscopic approach of Ref.~\cite{shahnazaryan2025polarons}, but aiming at general description polaron effects valid for arbitrary coupling strength. We restate the formalism in a way convenient for this purpose, carefully analyze the choice of relevant parameters, and introduce the effective dimensionless coupling constant that provides a proper 2D generalization of the famous Fr\"olich constant. We then examine the polaron in both weak and strong electron-phonon coupling regimes, and generalize the Feynman path-integral approach, which is known for its remarkable accuracy in interpolating between these two limits. By analyzing the polaron binding energy and effective mass within the Feynman all-coupling formalism, we clarify the role of dimensionality and screening in formation of 2D polaron states.

The paper is organized as follows. To make the paper self-contained and to restate the theory in a form suitable for a subsequent application of the path integral technique, in Sec.~II we derive the classical Lagrangian and Hamiltonian describing TO and LO phonons in 2D polar crystals, obtain the dispersion relation of LO phonons and discuss its interpretation as a non-local Lyddane-Sachs-Teller relation. The electron-phonon coupling terms, both in the Lagrangian and Hamiltonian settings, are derived in Sec.~III. Here, we also discuss relevant dimensionless parameters and introduce the effective coupling constant that measures the strength of electron-LO phonon interaction in 2D crystals. In Sec.~VI we consider the polaron ground state in the weak- and strong-coupling limits. In Sec.~V, the Feynman variational path-integral approach is adopted for our case of 2D polar crystals. We show how it interpolates between the limits of weak and strong coupling, providing quantitative estimates of both the polaron binding energy and effective mass. The results of numerical calculations within the Feynman approach are presented in Sec.~VI. We discuss applications to specific materials and provide a general analysis of the crossover between the weak- and strong-coupling regimes. The main results of this work are summarized in  the Conclusion with a discussion of the broader implications for transport and optical properties of 2D polar materials.

%%%%%%%%%%%%%%%%%%%%%%%%%%%%%%%%%%%%%%%%%%%%%%%%%%%%%%%%%%%%%%%%%%%%%%
\section{Macroscopic theory of 2D optical phonons}

Let us consider a generic 2D polar crystal containing sublattices of positive and negative ions. The macroscopic theory of long wavelength optical phonons in the 3D case dates back to the works of Born and Huang Kun \cite{BornHuang1954}, and by now constitutes a classical part of solid state physics (see, for example Ref. \cite{kittel1987quantum}). A proper 2D generalization of this macroscopic approach has recently been developed in our previous work~\cite{shahnazaryan2025polarons}. Here, to make the presentation self-contained, we present the main results of that theory and adopt its formulation for the specific purpose of this paper.

\subsection{Construction of Lagrangian density}

Our main goal is to describe long wavelength dynamics of the relative displacement field $\mathbf{S}$,
\begin{align}
\mathbf{S} = \mathbf{U}_+ - \mathbf{U}_-,
\end{align}
where $\mathbf{U}_+$ and $\mathbf{U}_-$ are the displacement vectors of positive and negative ions, respectively. For further convenience, we define the mass-weighted displacement field $\bm{\xi}$ as follows,
\begin{align}
\bm{\xi} = \mathbf{S}\sqrt{n M},
\end{align}
with $n = N/A$ being the surface density of unit cells, where $A$ is the normalization area, and $M$ the reduced ionic mass,
\begin{align}
M = \frac{m_+\,m_-}{m_+ + m_-}.
\end{align}
In the long wavelength limit, that is, assuming that the displacement varies on the scale much large than the interatomic distance and the thickness of the 2D layer, we write the Lagrangian density of the system as
\begin{align} \label{L-basic}
\mathcal{L} = \biggl( \frac{1}{2} \dot{\bm{\xi}}^2 - W^{\rm 2D} \biggr)\,\delta(z) 
+ \epsilon_0 \varepsilon \,\frac{\mathbf{E}^2}{2},
\end{align}
where the first two terms (in parentheses) represent the kinetic and potential energy contributions (per unit area) localized in the plane that hosts the lattice ($z=0$). The last term is the usual electromagnetic energy density in the surrounding medium with the dielectric constant $\varepsilon$ ($\epsilon_0$ stands for the vacuum permittivity). It should be emphasized that while the moving ions are bound to the 2D plane, they produce an electric field throughout the 3D space, which is crucial for a correct description of the dielectric screening in 2D materials \cite{cudazzo2011dielectric}. 
The potential energy density $W^{\rm 2D}$ in Eq.~\eqref{L-basic} accounts for the elastic energy of ionic displacements and their interaction with the electric field $\mathbf{E}$. Assuming for simplicity a macroscopic isotropy of the 2D layer, we write \footnote{An anisotropic generalization can be worked out straightforwardly.} 
\begin{align}\label{W-basic}
W^{\rm 2D}\! =\! \frac{1}{2} \omega_{\rm t}^2 \bm{\xi}^2\! -\! \gamma\,\bm{\xi}\cdot \mathbf{E}(\mathbf{r},z=0)\!
-\! \frac{\alpha_\infty}{2}\,\mathbf{E}^2(\mathbf{r},z=0),
\end{align}
which is analogous to the corresponding expression in the 3D case \cite{kittel1987quantum}. Here, $\omega_{\rm t}$ is the transverse optical (TO) phonon frequency, $\gamma$ parametrizes a coupling between $\bm{\xi}$ and $\mathbf{E}$, and $\alpha_\infty$ is the high-frequency (optical) 2D polarizability of a freestanding layer, which accounts for the electronic contribution to the electric response. By definition, the resulting two-dimensional polarization field is
\begin{align}
\mathbf{P}^{\rm 2D} = -\frac{\partial W^{\rm 2D}}{\partial \mathbf{E}} 
= \gamma\,\bm{\xi} + \alpha_\infty\, \mathbf{E}.
\end{align}
Taking the derivative with respect to $\bm{\xi}$ gives
\begin{align}
\frac{\partial W^{\rm 2D}}{\partial \bm{\xi}} = \omega_{\rm t}^2\,\bm{\xi} - \gamma\,\mathbf{E}.
\end{align}
In equilibrium (static limit $\omega=0$), we must have
\begin{align}
\omega_{\rm t}^2\,\bm{\xi}_0 - \gamma\,\mathbf{E} = 0,
\end{align}
which leads to the following relations,
\begin{align}
&\bm{\xi}_0 = \frac{\gamma}{\omega_{\rm t}^2}\,\mathbf{E}, \quad \mathbf{P}^{\rm 2D}_0 = \gamma\,\bm{\xi}_0 + \alpha_\infty\,\mathbf{E}= \alpha_0\,\mathbf{E}.
\end{align}
Thus, the full static 2D polarizability is identified as
\begin{align}
\alpha_0 = \alpha_\infty + \frac{\gamma^2}{\omega_{\rm t}^2},
\end{align}
which eventually allows us to express the coupling parameter $\gamma$ in Eq.~\eqref{W-basic} in terms of the experimentally accessible parameters, the static and optical 2D polarizabilities, and the frequency of TO phonons,
\begin{equation}
\gamma = \omega_{\rm t}\,\sqrt{\alpha_0 - \alpha_\infty}. \label{gamma} 
\end{equation}

\subsection{Dispersion of LO phonons in 2D}

From the Lagrangian, the equations of motion for the mass-weighted displacement field read as follows
\begin{align}
\ddot{\bm{\xi}} = -\,\omega_{\rm t}^2\,\bm{\xi} + \gamma\, \mathbf{E}(\mathbf{r},z=0).
\end{align}
Gauss's law, which is obtained by defining $\mathbf{E}=-\nabla\varphi$ into the Lagrangian Eq.~\eqref{L-basic} and minimizing it with respect to the electrostatic potential $\varphi$, connects the electric field divergence to the induced charge density via
\begin{align}
\nabla\cdot \mathbf{E}(\mathbf{r},z) = \frac{\rho}{\varepsilon\,\epsilon_0} 
\equiv -\,\frac{1}{\varepsilon\,\epsilon_0}\,\nabla\cdot \mathbf{P}^{\rm 2D}\,\delta(z).
\end{align}
Hence, once $\bm{\xi}$ is specified, the electrostatic potential can be determined.

By substituting $\mathbf{E}=-\nabla\varphi$ into the equations of motion, we have
\begin{align}
\label{eq:xit_new}
\ddot{\bm{\xi}}\,\delta(z) &= \bigl(-\omega_{\rm t}^2\bm{\xi}-\gamma\nabla\varphi\!\mid_{z=0}\bigr)\,\delta(z),\\
\nabla^{2}\varphi(\mathbf r,z) &= \frac{1}{\varepsilon\epsilon_{0}}\,
     \nabla\!\cdot\!\Bigl(\gamma\bm{\xi}-\alpha_\infty\nabla\varphi\!\mid_{z=0}\Bigr)\,\delta(z).
\end{align}
Assuming plane-wave normal mode solutions with in-plane wave vector $\mathbf{q}$, we perform the Fourier transform and obtain the following result for the Fourier component of the electrostatic potential,
\begin{align}\label{eq:four_pot}
\varphi_{\mathbf{q},k_z} &= - \frac{1}{\varepsilon \epsilon_0} \frac{1 }{q^2 + k_z^2} \left( i \gamma \mathbf{q} \cdot \bm{\xi}_{\mathbf{q}} + \alpha_\infty q^2 \varphi_{\mathbf{q}}^{\rm 2D} \right).
\end{align}
Here, the following conventions for the Fourier representation are used. In the plane, we define,  
\begin{align}
\bm{\xi}_{\mathbf q} &=\int {\rm d}^{2}r\,\bm{\xi}(\mathbf r) \frac{e^{-i\mathbf q\cdot\mathbf r}}{\sqrt{A}},\quad 
\bm{\xi}(\mathbf r) =\sum_{\mathbf q}\bm{\xi}_{\mathbf q}\frac{e^{i\mathbf q\cdot\mathbf r}}{\sqrt{A}},
\end{align}
while in $z$-direction the Fourier transform reads
\begin{align}
g_{k_{z}} &=\int\limits_{-\infty}^{\infty}{\rm d}z\,e^{-ik_{z}z}g(z),\quad
g(z) =\int\limits_{-\infty}^{\infty}\frac{{\rm d}k_{z}}{2\pi}\,e^{ik_{z}z}g_{k_{z}},
\end{align}
and also the in-plane slice of the potential was introduced,
\begin{align}
\varphi_{\mathbf q}^{\rm 2D} &=\frac{1}{2\pi}\int\limits_{-\infty}^{\infty} {\rm d}k_{z}\,\varphi_{\mathbf q,k_{z}} \equiv \varphi_\mathbf{q}(z=0).
\end{align}
The integration of Eq.~\eqref{eq:four_pot} over $k_{z}$ gives,
\begin{align}
\label{eq:phi2D_new}
\varphi_{\mathbf q}^{\rm 2D}
   = -\frac{\gamma\,\xi_{\mathbf q}^{l}}
            {2\varepsilon\epsilon_{0}\!
             \bigl[1+\alpha_\infty q/(2\varepsilon\varepsilon_{0})\bigr]},
\end{align}
where we defined the longitudinal component of the displacement as $\xi_{\mathbf q}^{\rm l} = i\mathbf q\cdot\bm{\xi}_{\mathbf q}/q$. 

Substituting $\nabla\varphi\!\mid_{z=0}=i\mathbf q\varphi_{\mathbf q}^{\rm 2D}$
into the Fourier form of Eq.~\eqref{eq:xit_new} we find  
\begin{align}
\omega^{2}\,\xi_{q,\omega}^{\rm l}
   = \omega_{\rm t}^{2}\,\xi_{q,\omega}^{\rm l}
      -\gamma\, q\,\varphi_{\mathbf q}^{\rm 2D}.
\end{align}
Finally, using Eq.~\eqref{eq:phi2D_new} and $\gamma^{2}=\omega_{\rm t}^{2}(\alpha_{0}-\alpha_\infty)$, we recover the dispersion relation for the LO phonon mode, obtained recently in Ref.~\cite{shahnazaryan2025polarons},
\begin{align} \label{omega-l}
\omega_{{\rm l},q}^{2}
   = \omega_{\rm t}^{2}\,
      \frac{1+\alpha_{0}q/(2\varepsilon\epsilon_{0})}
           {1+\alpha_\infty q/(2\varepsilon\epsilon_{0})}
   =  \omega_{\rm t}^{2}\,
      \frac{1+r_{0}q}{1+r_{\infty}q},
\end{align}
with 2D screening lengths,
$r_{\infty} = \alpha_\infty/(2\epsilon_{0}\varepsilon)$,
$r_{0}      = \alpha_{0}/(2\epsilon_{0}\varepsilon)$,
related to optical and static polarizabilities \footnote{We notice that the polarizabilities $\alpha_0$ and $\alpha_\infty$ refer to a freestanding 2D layer, whereas the corresponding screening length are defined for the layer in the dielectric environment. That is, to simplify formulas, the dielectric constant $\varepsilon$ of the environment is absorbed into the definition of $r_0$ and $r_\infty$.}. The dispersion relation of Eq.~\eqref{omega-l} agrees with earlier heuristic considerations supported by first-principle calculations \cite{sohier2016two,sohier2017breakdown,ponce2023long}. Recently, it has been observed experimentally \cite{Li2024experiment}.

The nontrivial dispersion of LO phonons in 2D polar insulators is a direct consequence of their nonlocal dielectric screening \cite{cudazzo2011dielectric}. An interesting observation is that Eq.~\eqref{omega-l} can be interpreted as a 2D form of the Lyddane-Sachs-Teller relation. In fact, the quantity $\varepsilon(q)=\varepsilon + \alpha q/2\epsilon_{0}$ plays a role of an effective dielectric constant in 2D, and therefore Eq.~\eqref{omega-l} can naturally be represented in the form,
\begin{align}
    \label{LST-relation}
    \omega_{{\rm l},q}^2 = \omega_{\rm t}^2\frac{\varepsilon_0(q)}{\varepsilon_\infty(q)}.
\end{align}

\section{Electron-phonon Lagrangian}

Having in mind the subsequent application to the path-integral formulation of the polaron problem, we now cast the phonon Lagrangian in the form of a set of independent normal modes.  All quantities in the following are expressed in the momentum
representation introduced in the previous section. That is, the in‐plane wave vectors, which count normal modes, assumed discrete (attributed to a large, but finite normalization area $A$), while the out‐of‐plane wave vectors $k_{z}$ are continuous and integration over $k_z$ is assumed everywhere.

\subsection{Normal mode phonon Lagrangian}

The original Lagrangian density of Eq.~\eqref{L-basic} explicitly reads as follows,
\begin{multline}
\mathcal{L}
=
\left(
\frac{1}{2}\,\dot{\bm{\xi}}^2 
-
\frac{\omega_{\rm t}^2}{2}\,\bm{\xi}^2 
+
\gamma\,\bm{\xi}\cdot \mathbf{E}(\mathbf{r},z=0)\right. \\ \label{L-phonon1}
\left.+ 
\frac{\alpha_\infty}{2}\,\mathbf{E}^2(\mathbf{r},z=0)
\right)\delta(z)
+
\epsilon_0\varepsilon\frac{\mathbf{E}^2}{2},
\end{multline}
resulting in the Lagrangian
\begin{align}
    L = \int \mathcal{L} {\rm d^2} r {\rm d}z .
\end{align}
In order to obtain the Lagrangian in terms of normal modes, we should substitute the electric field $\mathbf{E}=-\,\nabla\varphi$ in the Fourier representation. Using
the relation of Eq.~\eqref{eq:four_pot}, we find the result,
\begin{multline}
\nabla\varphi(\mathbf r,z)=-
\frac{i\gamma}{\epsilon_{0}\varepsilon\sqrt{A}}
\sum_{\mathbf q}\!
\frac{q\,\xi^{\rm l}_{\mathbf q}\,e^{i\mathbf q\cdot\mathbf r}}
     {1+r_\infty q}\\
\times\int\limits_{-\infty}^{\infty}\!\frac{{\rm d}k_{z}}{2\pi}\,
  \frac{\mathbf q+\hat{\mathbf z}k_{z}}
       {q^{2}+k_{z}^{2}}\,
  e^{ik_{z}z}.
\end{multline}
which, at $z=0$ (the location of the 2D crystal) takes the form,
\begin{align}
\nabla\varphi(\mathbf r,0)
  &= -\frac{i\gamma}{2\epsilon_{0}\varepsilon\sqrt{A}}
     \sum_{\mathbf q}
     \frac{\mathbf{q}}{1+r_\infty q}\,
     \xi^{\rm l}_{\mathbf q}\,
     e^{i\mathbf q\cdot\mathbf r}.                                  
\label{eq:nablaphi0}
\end{align}
By separating the longitudinal and transverse components of the displacement as $\bm{\xi}_{\mathbf q}=-i\frac{\mathbf{q}}{q}
\xi^{\rm l}_{\mathbf q}+ \bm{\xi}^{\perp}_{\mathbf q}$, and using
Eq.~\eqref{eq:nablaphi0} we obtain,
\begin{align}
&\frac12\!\int\!\dot{\bm{\xi}}^{2}\delta(z) {\rm d}^{2}r\,{\rm d}z
  =\frac12\sum_{\mathbf q}\!\bigl(|\dot{\xi}^{\rm l}_{\mathbf q}|^{2}
    +|\dot{\bm{\xi}}^{\perp}_{\mathbf q}|^{2}\bigr),                    \\
&\frac{\omega_{\rm t}^{2}}{2}\!\int\!\bm{\xi}^{2}\delta(z) {\rm d}^{2}r\, {\rm d}z
  =\frac{\omega_{\rm t}^{2}}{2}\sum_{\mathbf q}\!
    \bigl(|\xi^{\rm l}_{\mathbf q}|^{2}
          +|\bm{\xi}^{\perp}_{\mathbf q}|^{2}\bigr),                   \\
&\gamma\!\int\!\bm{\xi}\!\cdot\!\nabla\varphi(\mathbf r,0)\delta(z) {\rm d}m ^{2}r\, {\rm d}z\nonumber\\
 &\qquad\qquad\qquad =-\omega_{\rm t}^{2}
     \sum_{\mathbf q}
     \frac{(r_{0}-r_{\infty})\,q}
          {1+r_{\infty}q}\,
     |\xi^{\rm l}_{\mathbf q}|^{2},                                    \\
&\frac{\alpha_\infty}{2}\!\int\![\nabla\varphi(\mathbf r,0)]^{2}
           \delta(z) {\rm d}^{2}r\, {\rm d}z\nonumber\\
  &\qquad\qquad\qquad=\frac{\omega_{\rm t}^{2}}{2}
     \sum_{\mathbf q}
     \frac{(r_{0}-r_{\infty})\,q\,r_{\infty}q}
          {(1+r_{\infty}q)^{2}}\,
     |\xi^{\rm l}_{\mathbf q}|^{2},                                    \\
&\frac{\varepsilon\epsilon_{0}}{2}\!\int\!(\nabla\varphi)^{2}
           {\rm d}^{2}r\,{\rm d}z
  =\frac{\omega_{\rm t}^{2}}{2}
     \sum_{\mathbf q}
     \frac{\varepsilon\,(r_{0}-r_{\infty})\,q}
          {(1+r_{\infty}q)^{2}}\,
     |\xi^{\rm l}_{\mathbf q}|^{2}.
\end{align}
By collecting all terms together in Eq.~\eqref{L-phonon1} we arrive at the following result for the Lagrangian of optical phonons,
\begin{align}
L &=L_{\text{TO}}+L_{\text{LO}},       
\end{align}
where, as expected, the contributions of the transverse and longitudinal modes read, 
\begin{align}
L_{\text{TO}}&=\frac12\sum_{\mathbf q}\!
  \Bigl(|\dot{\bm{\xi}}^{\perp}_{\mathbf q}|^{2}
        -\omega_{\rm t}^{2}|\bm{\xi}^{\perp}_{\mathbf q}|^{2}\Bigr),     \\
L_{\text{LO}}&=\frac12\sum_{\mathbf q}\!
  \Bigl(|\dot{\xi}^{\rm l}_{\mathbf q}|^{2}
        -\omega_{{\rm l},q}^{2}|\xi^{\rm l}_{\mathbf q}|^{2}\Bigr).
\end{align}
Apparently, each mode $\xi^{\rm l}_{\mathbf q}$ and
$\bm{\xi}^{\perp}_{\mathbf q}$ behaves as an independent harmonic
oscillator with natural frequencies $\omega_{{\rm l},q}$ and
$\omega_{\rm t}$, respectively.

\subsection{Electron–phonon interaction in the Lagrangian framework}

Finally, to include interaction with a charge carrier (electron or hole), one notices that a charge $-e$ located at $\mathbf r_{e}$ couples to LO optical phonons via the electrostatic potential, yielding the following correction to the Lagrangian, 
\begin{align}
\Delta L \;=
e
\int d^2r
\,
\delta(\mathbf{r}-\mathbf{r}_e)
\,
\varphi(\mathbf{r},0)=\;e\,\varphi(\mathbf r_{e},0).
\end{align}
Using $\varphi(\mathbf r,0)=A^{-1/2}\sum_{\mathbf q}  \varphi^{\text{2D}}_{\mathbf q}\,e^{i\mathbf q\cdot\mathbf r}$
together with Eq.~\eqref{eq:phi2D_new}, we obtain
\begin{align}
\varphi(\mathbf r,0)=
-\frac{\gamma}{2\varepsilon\epsilon_{0}\sqrt{A}}
\sum_{\mathbf q}
\frac{\xi^{\rm l}_{\mathbf q}}
     {1+r_\infty q }\,
e^{i\mathbf q\cdot\mathbf r},
\end{align}
hence
\begin{multline}
\Delta L=-
\frac{e\gamma}{2\epsilon_{0}\varepsilon\sqrt{A}}
\sum_{\mathbf q}
\frac{\xi^{\rm l}_{\mathbf q}\,e^{i\mathbf q\cdot\mathbf r_{e}}}
     {1+r_\infty q}\\
=
\frac12\sum_{\mathbf q}g_{q}
\Bigl(e^{i\mathbf q\cdot\mathbf r_{e}}\xi^{\rm l}_{\mathbf q}
     +e^{-i\mathbf q\cdot\mathbf r_{e}}\xi^{\rm l}_{-\mathbf q}\Bigr),
\end{multline}
with a real coupling constant,
\begin{align}
g_{q}=
-\frac{e\gamma}
     {2\epsilon_{0}\varepsilon\sqrt{A}\,
      [1+r_\infty q]}.
\end{align}
The overall electron-LO phonon Lagrangian is then
\begin{multline}\label{eqn:final_lag}
L_{\text{e-ph}}=
\frac{m_{e}\dot{\mathbf r}_{e}^{2}}{2}
+\frac12\sum_{\mathbf q}
\bigl(|\dot{\xi}^{\rm l}_{\mathbf q}|^{2}
      -\omega_{{\rm l},q}^{2}|\xi^{\rm l}_{\mathbf q}|^{2}\bigr)\\
+\frac12\sum_{\mathbf q}g_{q}
\Bigl(e^{i\mathbf q\cdot\mathbf r_{e}}\xi^{\rm l}_{\mathbf q}
     +e^{-i\mathbf q\cdot\mathbf r_{e}}\xi^{\rm l}_{-\mathbf q}\Bigr).
\end{multline}
The TO phonons do not produce a long-range electric field, do not couple to the electrons electrostatically, and are irrelevant for the Fr\"ohlich polaron problem. We therefore ignore the $L_{\rm TO}$ part of the Lagrangian in the following consideration. In Sec.~V the Lagrangian~\eqref{eqn:final_lag} will be used within the path integral framework to develop a 2D generalization of the Feynman's all-couplings variational theory of the Fr\"ohlich polaron.

\subsection{Hamiltonian formulation}

It is also useful to restate the electron-LO phonon problem in the Hamiltonian formalism. In particular, we need it for the next section, where the weak-coupling limit will be analyzed within the perturbation theory. 

Firstly, within the standard canonical quantization, we express the coordinate and momentum operators for LO modes in terms of the corresponding phonon creation and annihilation operators. For each in–plane momentum $\mathbf q$, we have:
\begin{align}
&\xi^{\rm l}_{\mathbf q}=
\sqrt{\frac{\hbar}{2\omega_{{\rm l},q}}}\,
\bigl(\hat{a}_{\mathbf q}+\hat{a}_{-\mathbf q}^{\dagger}\bigr),\\
&\hat{\pi}_{\mathbf q}=\dot{\xi}^{\rm l}_{\mathbf q}=
-i\sqrt{\frac{\hbar\omega_{{\rm l},q}}{2}}\,
\bigl(\hat{a}_{\mathbf q}-\hat{a}_{-\mathbf q}^{\dagger}\bigr),
\end{align}
with $[\hat{a}_{\mathbf q},\hat{a}_{\mathbf q'}^{\dagger}]
     =\delta_{\mathbf q,\mathbf q'}$.
The phonon Hamiltonian is then obtained via the Legendre transform,
\begin{align}
\hat{H}_{\rm ph}=\sum_{\mathbf q}
\hbar\omega_{{\rm l},q}\,\hat{a}^{\dagger}_{\mathbf q}\hat{a}_{\mathbf q}.
\end{align}
By expressing $\xi^{\rm l}_{\mathbf q}$ in $\Delta L$ in terms of
$\hat{a}_{\mathbf q},\hat{a}_{\mathbf q}^{\dagger}$ we get the 2D
Fr\"{o}hlich interaction, and eventually the electron-phonon Hamiltonian,
\begin{multline}
\hat{H}_{\text{e-ph}}=
\frac{\hat{\mathbf p}_{e}^{2}}{2m_{e}}
+\hat{H}_{\rm ph}
\\+\sum_{\mathbf q}
V_{q}\Bigl(
      \,e^{i\mathbf q\cdot\hat{\mathbf r}_{e}}\hat{a}_{\mathbf q}
     +\,e^{-i\mathbf q\cdot\hat{\mathbf r}_{e}}
              \hat{a}^{\dagger}_{\mathbf q}
\Bigr), \label{H-e-ph}
\end{multline}
with the following coupling amplitude,
\begin{align} \label{V-original}
V_{q}=-g_{q}\sqrt{\frac{\hbar}{2\omega_{{\rm l},q}}}
     = \frac{e  \omega_{\rm t} }{2 \epsilon_0\varepsilon}
    \sqrt{ \frac{\hbar (\alpha_0 - \alpha_\infty) }{2A \omega_{{\rm l},k} } } \frac{ 1}{1 +  r_\infty k}.
\end{align}
This Hamiltonian constitutes the starting point for the perturbative and analyses that follow, while the Lagrangian form above is convenient for path-integral approaches.

Interestingly, the coupling amplitude $V_q$ of Eq.~\eqref{V-original} can be represented in a form that is similar to the standard Fr\"ohlich interaction with 3D bulk LO phonons \cite{frohlich1950xx,lee1953motion, kittel1987quantum}, and with dispersionless 2D surface phonons at surfaces of 3D polar crystals \cite{Sak1972,xiaoguang1985exact,peeters1986ground,hahn2018diagrammatic,li2018optical}. To this end, we introduce an effective 
\emph{non-local} dimensionless Fr\"ohlich constant $\alpha(q)$:
\begin{align} 
\alpha(q)=\dfrac{e^{2}}{4\pi\epsilon_0}\dfrac{1}{\hbar\omega_{\textup{l},q}}\sqrt{\frac{m_e\omega_{\textup{l},q}}{2\hbar}}
          \left[\dfrac{1}{\varepsilon_{\infty}(q)}-\dfrac{1}{\varepsilon_{0}(q)}\right], \label{eqn:fro_alpha}
\end{align}
where $\varepsilon_{0,\infty}(q)=\varepsilon(1+r_{0,\infty}q)$ are the effective 2D dielectric "constants", and the frequency of LO phonons satisfies the \emph{non-local} Lyddane-Sachs-Teller relation of Eq.~\eqref{LST-relation},
\begin{align}
 \omega_{{\rm l},q}
   =\omega_{\rm t}\sqrt{\dfrac{\varepsilon_{0}(q)}{\varepsilon_{\infty}(q)}}  
\end{align}
Then, the amplitude of 2D electron-phonon coupling in Eq.~\eqref{V-original} can be represented as follows,  
\begin{align} \label{V-alpha}
V_{q}=\!\left[\dfrac{\pi\alpha(q)\hbar}{Aq}\;
               \hbar\omega_{{\rm l},q}\sqrt{\dfrac{2\hbar\omega_{{\rm l},q}}{m}}\right]^{1/2},
\end{align}
which is exactly the form commonly used for the coupling to quasi-2D surface phonons \cite{Sak1972,xiaoguang1985exact,peeters1986ground,hahn2018diagrammatic,li2018optical}. However, in the present case of genuine 2D phonons, both the Fr\"{o}hlich coupling constant, and the frequency of LO phonons become $q$-dependent. This dependence, reflecting the non-locality of dielectric screening in 2D, is a unique feature of the 2D polaron problem, which makes it substantially different from corresponding problems both in the bulk and on the surface of 3D polar materials. 

\subsection{Characterization of the coupling strength in 2D.}

In the cases of 3D crystals, both in the bulk and on the surface, the polaron problem is fully determined by a single dimensionless parameter, the coupling constant $\alpha$. In 2D, the effective Fr\"ohlich constant Eq.~\eqref{eqn:fro_alpha} acquires the wave vector dependence, and the very characterization of the coupling strength becomes nontrivial.

In general, the electron-phonon problem of Eqs.~\eqref{H-e-ph}-\eqref{V-original} contains four length scales, the screening lengths, $r_\infty$ and $r_0$, the effective Bohr radius $a_{\rm B}$, and the oscillator length related to the frequency of TO phonons,
\begin{align}
 a_{\rm B} =\dfrac{4\pi\epsilon_0\varepsilon\hbar^2}{e^2m_e}, \quad   
 r_{\rm t}=\sqrt{\frac{\hbar}{2m_{e}\omega_{\rm t}}}. 
\end{align}
Three independent ratios of these lengths determine three independent dimensionless parameters that fully characterize the problem of 2D Fr\"ohlich polaron. 

Our choice of these tree parameters is based on the following observation. By inspecting the $q$-dependence of the coupling constant in Eq.~\eqref{eqn:fro_alpha} we notice that $\alpha(q)$ is bounded from above.  Firstly, the frequency of LO phonons always satisfies the inequality $\omega_{{\rm l},q}\ge\omega_{\rm t}$. Secondly, the factor $\varepsilon_\infty^{-1}(q)-\varepsilon_\infty^{-1}(q)$ vanishes in the limits $q\to 0$ and $q\to\infty$, and has a single maximum at $q=q_{\rm m}=1/\sqrt{r_0 r_\infty}$. Therefore, by replacing $\omega_{{\rm l},q}\to \omega_{\rm t}$, and evaluating $\varepsilon_\infty^{-1}(q)-\varepsilon_\infty^{-1}(q)$ at $q=q_{\rm m}$, we find for $\alpha(q)$,
\begin{align}
    \label{alpha-m}
\alpha(q) < \alpha_{\rm m} = \frac{r_{\rm t}}{a_{\rm B}}\frac{\sqrt{r_0}-\sqrt{r_\infty}}{\sqrt{r_0}+\sqrt{r_\infty}}.    
\end{align}
In the following, we choose the value of $\alpha_{\rm m}$ as the first dimensionless parameter characterizing the system, while the other two are defined by the ratios,   
\begin{align}
&\sigma_{0}=\frac{r_{0}}{r_{\infty}}>1,\quad \sigma_{\rm t}=\frac{r_{\rm t}}{r_{\infty}}.
\end{align}

On physical grounds, it is natural to expect that $\alpha_{\rm m}$ indeed provides a proper measure of the coupling strength, which generalized the notion of Fr\"ohlich constant to the case of 2D polar materials 
\footnote{It is worth emphasizing that $\alpha_{\rm m}$ is the {upper bound} of the function $\alpha(\kappa)$ rather than its true maximum. The exact maximizer of $\alpha(\kappa)$ has a cumbersome closed form, whereas the maximizer of $\alpha(\kappa) [(1+ r_0 \kappa) / (1+r_\infty \kappa)]^{1/4} $,
which equals $\alpha_{\rm m}$, is both simple and physically transparent.}. This expectation is confirmed in the following sections by explicitly analyzing the polaron problem. We will see that the polaron energy depends most strongly on $\alpha_{\rm m}$ with the value of $\alpha_{\rm m}\sim 5$ separating the weak- and strong-coupling regimes. Moreover, for any value of $\alpha_{\rm m}$, the dependency on the other parameters is moderate.

\section{Polaron problem in weak and strong coupling Limits}
\subsection{Weak-coupling regime: Second-order perturbation theory}

In the limit $\alpha_{\rm m}\ll 1$, we adopt the perturbation theory, assuming the unperturbed ground state, which consists of an electron with zero momentum and the phonon vacuum,
$\lvert 0\rangle=\lvert\mathbf P{=}0\rangle\!\otimes\!\lvert 0\rangle_{\rm ph}$.
Because the interaction vertex
$V_{q}(e^{i\mathbf q\cdot\hat{\mathbf r}_{e}}\hat a_{\mathbf q}
      -e^{-i\mathbf q\cdot\hat{\mathbf r}_{e}}\hat a_{\mathbf q}^{\dagger})$
changes the phonon number by $\pm1$,  
the first-order correction to the energy vanishes.  
The leading energy shift appears in the second order,
\begin{align}
\Delta E_{0}=-
\sum_{n\neq 0}
\frac{|\langle n|\hat H_{\text{int}}|0\rangle|^{2}}
     {E_{n}^{(0)}-E_{0}^{(0)}},
\end{align}
where the intermediate states $\lvert n\rangle$
contain one LO phonon with the wave vector $\mathbf q$
and the electron recoiling with momentum $-\mathbf q$.
Evaluating the matrix elements we get the following result,
\begin{equation}
\Delta E_{0}=-
\sum_{\mathbf q}
\frac{|V_{q}|^{2}}
     {\dfrac{\hbar^{2}q^{2}}{2m_{e}}+\hbar\omega_{{\rm l},q}}\;.
\label{eq:DE0-sum}
\end{equation}  
After replacing summation by integration $\sum_{\mathbf q}\to \frac{A}{(2\pi)^{2}}\int {\rm d}^{2}q$, using Eq.~\eqref{V-alpha} to express $V_q$ in terms of $\alpha(q)$, the perturbative correction to the energy reduces to the following form,
\begin{align}
\Delta E_{0}=-
\hbar\int\limits_{0}^{\infty}{\rm d}q\,
\frac{\alpha(q)\;\hbar\omega_{{\rm l},q}}
     {\displaystyle 
      \frac{\hbar^{2}q^{2}}{2m_{e}}+\hbar\omega_{\mathrm l,q}}
      \sqrt{\frac{\hbar\omega_{\mathrm l,q}}{2m_{e}}} 
\label{eq:DE0-sum-alpha}
\end{align}
where we performed a trivial angular integration. Finally, we change the integration variable $p= r_\infty \,q$ and express the polaron energy shift in terms of dimensionless parameters introduced in Sec.~III~D,
\begin{align}
    E_0=\Delta E_0 = -\alpha_{\rm m} \hbar\omega_{\rm t} {\mathcal I}_0(\sigma_0,\sigma_{\rm t}),
    \label{eq:final-DeltaE0-weak}
\end{align}
where the dimensionless integral coefficient reads
\begin{align}\label{eqn:int_wc_dimless}
{\mathcal I}_0(\sigma_0,\sigma_{\rm t})
= \int\limits_{0}^{\infty} 
\frac{\sigma_{\rm t}(\sqrt{\sigma_0}+1)^2 p {\rm d}p}{(1+p)(1+\sigma_0 p)(\sigma_{\rm t}^2p^2/\Omega_p+1)}
\end{align}
with $\Omega_p=\sqrt{(1+\sigma_0p)/(1+p)}$ being the dimensionless frequency of LO phonons. 

It is clear from Eq.~\eqref{eq:final-DeltaE0-weak} that the parameter $\alpha_{\rm m}$ indeed plays the
same role as the coupling constant in the 3D Fr\"{o}hlich
problem. The remaining integral coefficient defined by
Eq.~\eqref{eqn:int_wc_dimless} turns out to be a well‑behaved dimensionless function of the order of unity in a wide range of its arguments. This allows us to roughly estimate the polaron binding energy as $E_{\rm b} \approx \alpha_{\rm m} \hbar\omega_{\rm t}$. 

\begin{figure}[t]
    \centering
    \includegraphics[width=\linewidth]{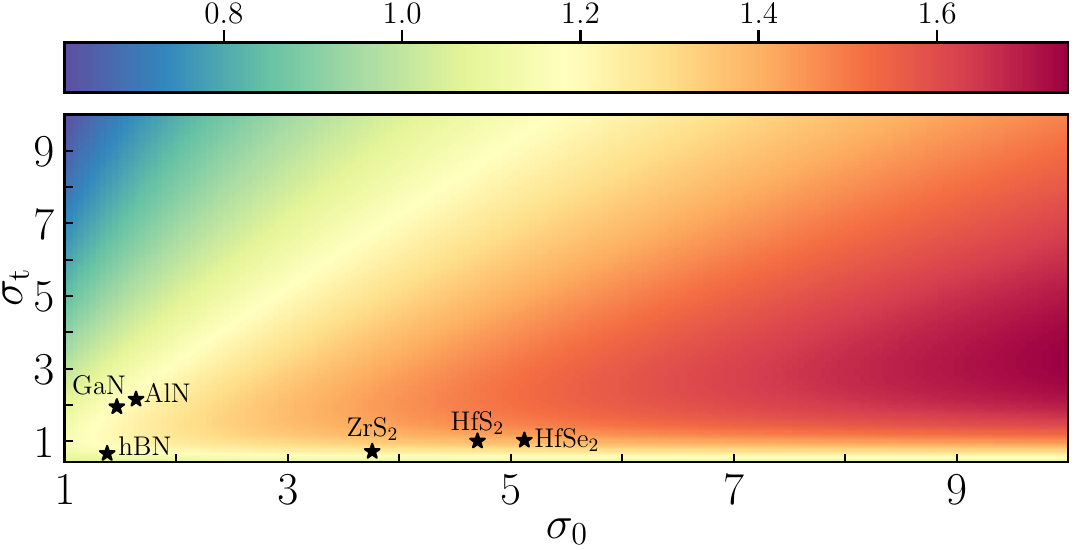}
    \caption{Colour map of the dimensionless weak‑coupling integral
    ${\mathcal I}_{0}(\sigma_{0},\sigma_{\rm t})$ defined in
    Eq.~\eqref{eqn:int_wc_dimless}.  
    The horizontal axis is the ratio
    $\sigma_{0}$ of static to high‑frequency screening
    lengths, the vertical axis is
    $\sigma_{\rm t}$.  
    Stars indicate the realistic parameter sets of hBN, GaN, AlN,
    ZrS$_2$, HfS$_2$, and HfSe$_2$.  
    The weak dependence on both variables (the colour scale spans only a
    factor of two) justifies treating $\alpha_{\rm m}$ as the principal
    measure of the electron–phonon coupling in the perturbative
    regime.}
    \label{fig:heat_map_wc}
\end{figure}

Figure~\ref{fig:heat_map_wc} displays a color map of ${\mathcal I}_{0}(\sigma_{0},\sigma_{\rm t})$ over a range of material parameters $\sigma_{0}$ and $\sigma_{\rm t}$ varying by an order of magnitude.
The stars indicate the positions of the specific 2D materials analyzed in Table~\ref{table:1}. Over the entire region, the function ${\mathcal I}_{0}(\sigma_{0},\sigma_{\rm t})$ varies only moderately, taking values of order unity, which confirms the interpretation of $\alpha_{\rm m}$ as the effective coupling constant controlling the overall magnitude of the polaron energy shift. We will see that this conclusion holds true for any coupling strength.

It is well known that for 3D polar crystals, the polaron energy shift calculated as the lowest-order perturbative correction is exactly reproduced within the so-called intermediate coupling variational theory by Lee, Low, and Pines (LLP) \cite{lee1953motion}. In 2D the situation is the same. In fact, the perturbative result of Eq.~\eqref{eq:final-DeltaE0-weak} exactly coincides with the polaron energy obtained in Ref.~\cite{shahnazaryan2025polarons} within the variational LLP approach. The variational derivation of the apparently perturbative result explains why the linear dependence of the polaron energy extends to quite large coupling both in 3D and in 2D, as we will see in Sec.~VI.

\subsection{Strong-Coupling Limit (Landau-Pekar Model)}
In the strong-coupling regime $\alpha_{\rm m}\gg 1$, the adiabatic approximation for the phonon subsystem is justified, and the polaron binding energy can be estimated using the Landau-Pekar energy functional (see, e.g., \cite{devreese2016fr}),
\begin{align}
    E_e^{\rm LP} = \Bigl\langle \psi\Bigl|\frac{\hat{p}_e^2}{2m_e}\Bigr|\psi\Bigr\rangle - \sum_{\mathbf k} \frac{|V_k|^2|g_k|^2}{\hbar\omega_{{\rm l},k}}.
    \label{eq:LP-E}
\end{align}
Here $|\psi(r)\rangle$ is the wave function of the self-trapped electron, and the corresponding form factor is defined as
\begin{align}
    g_k = \Bigl\langle \psi\Bigl|e^{i\mathbf{k}\cdot\mathbf{r}}\Bigr|\psi\Bigr\rangle.
\end{align}
\subsubsection{Hydrogen-like trial wavefunction}
For the simplest and frequently used hydrogenic trial wave function \cite{sio2023polarons} in 2D,
\begin{align}
    \psi_{\rm H} = \sqrt{\frac{2}{\pi}}\frac{1}{a}\,e^{-r/a},
    \label{eq:LP-trial-psi}
\end{align}
the kinetic energy is
\begin{align}
    \Bigl\langle \psi_{\rm H}\Bigl|\frac{\hat{p}_e^2}{2m_e}\Bigr|\psi_{\rm H}\Bigr\rangle = \frac{\hbar^2}{2m_ea^2},
    \label{eq:LP-kinetic}
\end{align}
and the density form factor reads,
\begin{align}
    g_q = \frac{1}{\Bigl(1+{(aq)^2}/{4}\Bigr)^{3/2}}.
    \label{eq:LP-gk}
\end{align}
Hence, the Landau–Pekar energy as a function of the variational parameter $a$ is as follows,
\begin{multline}
E_{\mathrm{LP}}^{\rm H}(a)=
\frac{\hbar^{2}}{2m_{e}a^{2}}
\\-\sum_{\mathbf q}
\frac{\pi\hbar\,\alpha(q)}{A\,q}\,
\sqrt{\frac{2\hbar\omega_{\mathrm l,q}}{m_{e}}}\,
\frac{1}{\bigl[\,1+(a q)^{2}/4\bigr]^{3}}\!,
\label{eq:LP-sum-alpha}
\end{multline}
where we used Eq.~\eqref{V-alpha} and expressed $V_q$ in terms of the non-local coupling constant $\alpha(q)$ and the frequency $\omega_{\mathrm l,q}$. Following the same steps as in the previous subsection, we convert sum to integral, express everything in terms of the three dimensionless parameters, and arrive at the final result,
\begin{align}
E_{\mathrm{LP}}^{\rm H}(a)=
\frac{\hbar^{2}}{2m_{e}a^{2}}
-\alpha_{\rm m} \hbar\omega_{\rm t} {\mathcal I}_{\rm LP}^{\rm H}(a,\sigma_0,\sigma_{\rm t}), \label{eq:final-DeltaE0-strong_h}
\end{align}
where the dimensionless integral in the second term reads
\begin{align}
{\mathcal I}_{\rm LP}^{\rm H}=\int\limits_{0}^{\infty}\dfrac{\sigma_{\rm t}(\sqrt{\sigma_0} + 1)^2\, {\rm d}p}{(1+p)(1+\sigma_0 p)\left[1+\left(p a/2r_\infty\right)^2\right]^3}.\label{eqn:int_sch_dimless}
\end{align}
The minimization of Eq.~\eqref{eq:LP-sum-alpha} (or equivalently Eq.~\eqref{eq:final-DeltaE0-strong_h}) with respect to $a$ yields in the strong‑coupling polaron binding energy.  The above Landau-Pekar type theory with the hydrogenic trial function can be directly compared with similar calculations in Ref.~\cite{sio2023polarons} based on a simplified model of electron-LO phonon coupling that neglects the phonon dispersion. As we have demonstrated recently, for sufficiently strong coupling, the correct treatment of the dispersion and the related nonlocality of the coupling constant becomes highly important (see the Supplemental Material in Ref. \cite{shahnazaryan2025polarons}). 

\subsubsection{Gaussian trial wavefunction}

Another popular variational ansatz, which typically gives lower energy, employs the normalized Gaussian that in 2D reads,   
\begin{align}
\psi_{\rm G}(\mathbf r)=\sqrt{\frac{1}{\pi a^{2}}}\;
                e^{-r^{2}/2a^{2}} .
\end{align}
The corresponding expectation value of the kinetic energy coincides with the one for hydrogenic ansatz,   
\begin{equation}
\Bigl\langle\psi_{\rm G}\Bigl|\frac{\hat p_{e}^{2}}{2m_{e}}\Bigr|\psi_{\rm G}\Bigr\rangle
      =\frac{\hbar^{2}}{2m_{e}a^{2}},
\label{eq:LP-Gauss-kin}
\end{equation}
while the form factor is now  
\begin{equation}
g_{q}(a)=e^{-a^{2}q^{2}/4}.
\label{eq:LP-Gauss-gq}
\end{equation}
Substituting \eqref{eq:LP-Gauss-kin} and \eqref{eq:LP-Gauss-gq} in the Landau–Pekar functional and expressing the coupling through the non-local Fr\"{o}hlich constant, we obtain the following variational energy,
\begin{equation}
E_{\mathrm{LP}}^{\rm G}(a)=
\frac{\hbar^{2}}{2m_{e}a^{2}}
-
\sum_{\mathbf q}
\frac{\pi\hbar\,\alpha(q)}{A\,q}\,
\sqrt{\frac{2\hbar\omega_{\mathrm l,q}}{m_{e}}}\;
e^{-a^{2}q^{2}/2},
\label{eq:LP-Gauss-sum}
\end{equation}
which eventually reduces to the form similar to Eq.~\eqref{eq:final-DeltaE0-strong_h},
\begin{align}
E_{\mathrm{LP}}^{\rm G}(a)=
\frac{\hbar^{2}}{2m_{e}a^{2}}
-\alpha_{\rm m} \hbar\omega_{\rm t} {\mathcal I}_{\rm LP}^{\rm G} (a,\sigma_0,\sigma_{\rm t}),\label{eq:LP-Gauss-int}
\end{align}
where for the Gaussian trial function, the dimensionless integral reads,
\begin{align}
{\mathcal I}_{\rm LP}^{\rm G}=\int\limits_{0}^{\infty}\frac{\sigma_{\rm t} (\sqrt{\sigma_0}+1)^2\,p}{(1+p)(1+\sigma_0 p)}\exp\left(-\dfrac{p^2a^2}{2 r_\infty^2}\right) {\rm d}p.\label{eqn:int_scg_dimless}
\end{align}
By minimizing the energy of Eq.~\eqref{eq:LP-Gauss-int} with respect to the variational parameter $a$ we get the Landau–Pekar polaron energy in the Gaussian trial class. As we have already mentioned, the Gaussian ansatz yields a better energy estimate compared to the hydrogenic one, but most importantly, it corresponds to the exact strong-coupling limit of the Feynman variational theory, which we consider in the following sections.

\section{Feynman Variational Solution of the Polaron Problem in 2D}

Feynman variational solution of the polaron problem, based on the path integral formalism \cite{feynman1955slow}, is known to provide a high-quality all-coupling theory that smoothly interpolates between week- and strong-coupling limits and perfectly agrees with essentially exact MC results both for the bulk 3D and the surface quasi-2D polarons  \cite{prokofev1998polaron,mishchenko2000diagrammatic,hahn2018diagrammatic}.  In the present section, we adopt this approach to the case of genuine  2D polarons. As the standard derivation \cite{feynman1955slow,Feynman-StatMech-book} of the corresponding variational problem assumes from the very beginning dispersionless phonons and a $q$-independent Fr\"ohlich coupling, its 2D generalization requires some care. Therefore, in the following, we carefully adopt the key steps of the Feynman derivation to our specific situation, avoiding unnecessary assumptions. This in particular allows us to keep the presentation fully self-contained.

\subsection{Effective action}

The first step is to cast the electron–phonon problem in imaginary time, starting from the real‑time Lagrangian of Eq.~ \eqref{eqn:final_lag} derived in  Sec.~III~B. The corresponding real-time action reads, 
\begin{multline}
S=\!\int\limits_{t_{\mathrm i}}^{t_{\mathrm f}}\!{\rm d}t
\Bigl[
  \tfrac{m_{e}}{2}\dot{\mathbf r}^{2}_{e}
 +\tfrac12\!\sum_{\mathbf q}\!
     \bigl(\dot{\xi}^{\rm l}_{\mathbf q}\dot{\xi}^{\rm l}_{-\mathbf q}
            -\omega_{{\rm l},q}^{2}\xi^{\rm l}_{\mathbf q}\xi^{\rm l}_{-\mathbf q}\bigr)\\
 +\tfrac12\!\sum_{\mathbf q}g_{q}
     \bigl(e^{i\mathbf q\!\cdot\!\mathbf r_{e}}\xi^{\rm l}_{\mathbf q}
          +e^{-i\mathbf q\!\cdot\!\mathbf r_{e}}\xi^{\rm l}_{-\mathbf q}\bigr)
\Bigr].
\label{eq:S-real}
\end{multline}
Introducing imaginary time $u$ by the standard complex rotation,  
$t\to -iu$, $\,{\rm d}t=-i\,{\rm d}u$, we generate the Euclidean action,
\begin{align}
S\;\longrightarrow\;S_{\mathrm E}=-\,iS
      =\!\int\limits_{0}^{\beta\hbar}\!{\rm d}u\,\mathcal L_{\mathrm E},
      \qquad \beta=\frac{1}{k_{\mathrm B}T},
\end{align}
where  $\mathcal L_{\mathrm E}$ is obtained from the real‑time Lagrangian by simply changing the sign in front of kinetic energy terms, $\dot{(\ \ )}^{\,2}\!\to-\,\dot{( \ \  )}^{\,2}_{u}$.

The phase factor $e^{\,iS}$ in the real-time propagator,  
after the above Wick rotation, becomes a Boltzmann weight in the path-integral representation of partition function,  
\begin{align}
\mathcal Z=\!\int\!\mathcal D\mathbf r_{e}(u)\,
            \mathcal D\xi^{\rm l}_{\mathbf q}(u)\;
            e^{-S_{\mathrm E}[\mathbf r_{e},\xi^{\rm l}]/\hbar} .
\end{align}
Performing the Gaussian integration over the phonon coordinates, one obtains an effective action that involves only the electron trajectory,
\begin{multline}
    S_{\mathrm{eff}}[\mathbf{r}_e] = \int\limits_{0}^{\beta\hbar} {\rm d}u\,\frac{m_e\,\dot{\mathbf{r}}_e^2(u)}{2} - \frac{1}{2\hbar}\int\limits_{0}^{\beta\hbar}\int\limits_{0}^{\beta\hbar}{\rm d}t\,{\rm d}s\,\\ 
    \times\sum_{\mathbf{k}}|V_k|^2\,e^{-\omega_{ {\rm l},k}|t-s|}e^{i\mathbf{k}\cdot\bigl[\mathbf{r}_e(t)-\mathbf{r}_e(s)\bigr]}.
\end{multline}
This action is used to define the upper bound of the polaron energy within the Feynman variational  principle. 

\subsection{The upper bound of the polaron ground state energy}
The starting point here is the following relation between the ground state energy $E$ and the effective action,
\begin{equation}
E
=
\lim_{\beta \to \infty}
\Bigl[
-\tfrac{1}{\beta}\,
\ln\,
e^{-\beta F} 
\Bigr],
\end{equation}
where $e^{-\beta F}=\mathcal{Z}$  is defined via the path integral,
\begin{align} 
e^{-\beta F} 
&= \int e^{- S_{\mathrm{eff}}[\mathbf{r}_e]/\hbar}\, \mathcal{D}\mathbf{r}_e.
\end{align}

The Feynman variational principle is formulated as follows. Let us define a trial (reference) action $S_0[\mathbf{r}_e]$, and the corresponding  ground state energy,
\begin{align}\label{E_0-def}
    &E_0 =
\lim_{\beta \to \infty}
\Bigl[
-\tfrac{1}{\beta}\,
\ln\,
e^{-\beta F_0} 
\Bigr], \\
&e^{-\beta F_0} 
= \int e^{- S_{0}[\mathbf{r}_e]/\hbar}\, \mathcal{D}\mathbf{r}_e.
\label{F0-def}
\end{align}
Then the upper bound for the exact energy of the ground state is given by the inequality \cite{Feynman-StatMech-book},
\begin{equation}
E\;\le\;
E_{0}+\lim_{\beta\to\infty}\frac{1}{\beta\hbar}\bigl\langle S_{\rm eff}-S_{0}\bigr\rangle_{0},
% \qquad
% E_{0}\equiv\lim_{\beta\to\infty}F_{0}.
\label{eq:var-bound-start}
\end{equation}
where the average with respect to the trial action $S_0$ is computed as follows, 
\begin{align}
\bigl\langle \mathcal{O} \bigr\rangle_{0}
=
\frac{\displaystyle \int \mathcal{O}\,\exp\bigl[-\,S_{0}/\hbar\bigr]\,\mathcal{D}\mathbf{r}_e}
{\displaystyle \int \exp\bigl[-\,S_{0}/\hbar\bigr]\,\mathcal{D}\mathbf{r}_e}.
\end{align}
The trial action should be defined with some variational freedom under the requirement that it does not restrict the electron motion too strongly, remaining a good variational model.

The Feynman classic approach employs a Gaussian trial action, which allows for an analytic treatment of the resulting path integrals and yields highly accurate results for the polaron energy, effective mass, and related characteristics. Specifically, the trial action is taken in the following form,
\begin{multline}
S_{0}=\int\limits_{0}^{\beta\hbar}\!\frac{m_{e}\dot{\mathbf r}_{e}^{2}(u)}{2}\, {\rm d}u
      +\frac{C}{2}\!
       \int\limits_{0}^{\beta\hbar}\!\!\int\limits_{0}^{\beta\hbar} {\rm d}u\,{\rm d}u' \\
         \times e^{-W|u-u'|}
         \bigl|\mathbf r_{e}(u)-\mathbf r_{e}(u')\bigr|^{2},
\end{multline}
where $[W]={\rm s}^{-1}$ and $ [C] = {\rm kg}\cdot {\rm s}^{-3}$ are two variational parameters. 

In order to evaluate the second term in the right-hand side in Eq.~ \eqref{eq:var-bound-start}  one needs two building blocks, first, the form-factor,
\begin{align} \label{form-factor}
        I(\mathbf k;u,v)=
        \bigl\langle
          e^{\,i\mathbf k\cdot[\mathbf r_{e}(u)-\mathbf r_{e}(u')]}
        \bigr\rangle_{0},
\end{align}
and, second, the displacement correlator, which can be found by differentiating $I(\mathbf k;u,u')$ with respect to $\mathbf{k}$:  
\begin{align}
        \bigl\langle|\mathbf r_{e}(u)-\mathbf r_{e}(v)|^{2}\bigr\rangle_{0}
        =-\nabla_{\mathbf k}^{2}
          I(\mathbf k;u,u')\bigl|_{\mathbf k=0},
\end{align}
Explicitly, the form factor $I(\mathbf k;u,u')$ in the limit $\beta\!\to\!\infty$ reads,
\begin{multline}
I(\mathbf k;u,u')=
\exp\!\Bigl[
  -\tfrac{k^{2}}{2}\Bigl(
      \tfrac{W^{2}\hbar}{V^{2}m_{e}}|u-u'|
     \\+\tfrac{4C\hbar}{WV^{3}m_{e}^2}
      \bigl[1-e^{-V|u-u'|}\bigr]\Bigr)
      \Bigr],
\label{eq:I-kuv}
\end{multline}
where $V=\sqrt{W^{2}+4C/(m_{e}W)}$.

Having the expression for $I(\mathbf k;u,u')$ at hand, we can evaluate the average $\langle S_{\rm eff}-S_{0}\rangle_{0}$.  Rewriting it in terms of $I(\mathbf k;u,u')$, one finds
\begin{multline}
\!\!\bigl\langle S_{\mathrm{eff}}-S_{0}\bigr\rangle_{0}
\!=\!
-\sum_{\mathbf k}
      \frac{|V_{k}|^{2}}{2\hbar}\!\!
      \int\limits_{0}^{\beta\hbar}\!\!\!{\rm d}u\!\!
      \int\limits_{0}^{\beta\hbar}\!\!{\rm d}u'\;
      e^{-\omega_{{\rm l},k}|u-u'|}
      I(\mathbf k;u,u')
\\-\frac{C}{2}
      \int\limits_{0}^{\beta\hbar}\!{\rm d}u
      \int\limits_{0}^{\beta\hbar}\!{\rm d}u'\;
      e^{-W|u-u'|}\,
      \Bigl[\nabla_{\mathbf k}^{2}I(\mathbf k;u,u')\Bigr]_{\mathbf k=0},
\label{eq:SminusS0-sum}
\end{multline}
We now evaluate three contributions entering the bound of Eq.~(\ref{eq:var-bound-start}): (i) the $C$–term originating from $S_{0}$, (ii) the reference ground‑state energy $E_{0}$, and (iii) the genuine electron–phonon contribution that contains
$|V_{k}|^{2}$.

%--------------------------------------------------
\paragraph*{(i)  The $C$–term.}

Using the definition of the displacement correlator and the explicit form of Eq.~(\ref{eq:I-kuv}) one finds in the limit $\beta\!\to\!\infty$,
\begin{align}
\frac{C}{2}&\!
\int\limits_{0}^{\beta\hbar}\!\!{\rm d}u
\int\limits_{0}^{\beta\hbar}\!\!{\rm d}u'\;
e^{-W|u-u'|}\left[-2\dfrac{{{\rm d}I(k,u,u') }}{{\rm d}k^2}\right]_{\mathbf{k}=0}\\
  & =\frac{2C\,\beta\hbar^{2}}{m_{e}WV} = \hbar^2\beta\frac{V^2-W^2}{2V},
\label{eq:C-term-final}
\end{align}
where we expressed the parameter $C$ in terms of $V$ and $W$ as $C=\tfrac{m_{e}W}{4}(V^{2}-W^{2})$.
%--------------------------------------------------
\paragraph*{(ii)  Trial ground‑state energy $E_{0}$.}

The free energy $F_{0}(C)$ of the Gaussian model follows from
$\partial_{C}F_{0}=
  \beta^{-1}\hbar^{-1}\langle\partial_{C}S_{0}\rangle_{0}$.
With the same Gaussian algebra that leads to Eq.~(\ref{eq:C-term-final}) we obtain $ \partial_{C}E_{0}=2\hbar/(m_{e}WV).$
Integrating this derivative and using the condition $E_{0}(C{=}0)=0$ gives the closed form of the reference energy,
\begin{align}
E_{0}= \hbar\,(V-W).
\label{eq:E0-result}
\end{align}

%--------------------------------------------------
\paragraph*{(iii)  Electron–phonon part.}
Only this contribution depends on the phonon dispersion and electron-phonon coupling, and therefore carries all information about specific features of the 2D system. Keeping the explicit coupling amplitude $|V_{k}|^{2}$ we define
\begin{align}
\mathcal F
&\equiv
\sum_{\mathbf k}\!\frac{|V_{k}|^{2}}{2\hbar}
      \int\limits_{0}^{\beta\hbar}\!{\rm d}u
      \int\limits_{0}^{\beta\hbar}\!{\rm d}u'\;
      e^{-\omega_{ {\rm l},k}|u-u'|}\,
      I(\mathbf k;u,v).
\end{align}
Using the identity
\begin{align}
\int\limits_{0}^{\beta\hbar}\!\!\int\limits_{0}^{\beta\hbar}
      {\rm d}u\,{\rm d}u'\;g(|u-u'|)
     \xrightarrow[\beta\to\infty]{}\;
  2\beta\hbar\int\limits_{0}^{\infty}\!{\rm d}u\,g(u)
\end{align}
and the fact that $I(\mathbf k;u,u')$ in Eq.~(\ref{eq:I-kuv}) depends on $|u-u'|$, we obtain
\begin{multline}
\mathcal F
= \beta
   \sum_{\mathbf k}\!
   |V_{k}|^{2}
   \int\limits_{0}^{\infty} {\rm d}u\;\exp\!\bigl[
     -\omega_{ {\rm l},k}u\bigr]\\
   \times\exp\!\Bigl[
     -\tfrac{k^{2}}{2}\Bigl(
          \tfrac{W^{2}\hbar}{V^{2}m_{e}}u
        + \tfrac{4C\hbar}{WV^{3}m_{e}^2}
          \bigl[1-e^{-Vu}\bigr]\Bigr)
       \Bigr].
\label{eq:I-sum-final}
\end{multline}
Collecting Eqs.~(\ref{eq:C-term-final}), (\ref{eq:E0-result}) and
(\ref{eq:I-sum-final}) in the right-hand side of Eq.~(\ref{eq:var-bound-start}) we arrive at the upper bound that depends on two dimensionless variational parameters $v\!=\!V/\omega_{\rm t}$ and $w\!=\!W/\omega_{\rm t}$,
\begin{align}
&\qquad E\le
\hbar\omega_{\rm t}\frac{(v-w)^{2}}{2v}\nonumber\\
  & -\int\limits_{0}^{\infty} {\rm d}u \sum_{\mathbf k}\frac{\pi \alpha(k)}{A \; k}\!
\hbar \omega_{{\rm l},k}\sqrt{\frac{2\hbar\omega_{{\rm l},k}}{m_e}}\;
     e^{-\omega_{ {\rm l},k}u}\nonumber\\
& \times\exp\!\Bigl[
     -\tfrac{\hbar k^{2}}{2m_e\omega_{\rm t}}\Bigl(
          \tfrac{w^{2}}{v^{2}}u\omega_{\rm t}
        + \tfrac{v^2-w^2}{v^3}
          \bigl[1-e^{-vu \omega_{\rm t} }\bigr]\Bigr)
       \Bigr].
\end{align}

After replacing the sum over momenta by  integral,
and a $q$-dependent rescaling of the time variable $u\!\to\!\tau\,\omega_{ {\rm l},q}$,
we arrive at the following final result for the variational upper bound,
\begin{align}
E\;\le\;
\hbar\omega_{\rm t}\,\frac{(v-w)^{2}}{2v}
-\hbar\omega_{\rm t}\,\alpha_{\rm m}\, 
{\mathcal I}_{\rm F}(\sigma_{0},\sigma_{\rm t};v,w),\label{eqn:upper_bound_int}
\end{align}
where we introduced the dimensionless function,
\begin{widetext}
\begin{align} \label{I-F}
{\mathcal I}_{\rm F}(\sigma_{0},\sigma_{\rm t};v,w)=
\int\limits_{0}^{\infty}\!
{\rm d}\tau
\int\limits_{0}^{\infty}\! {\rm d}p\;\frac{\sigma_{\rm t}(\sqrt{\sigma_0}+1)^2\,p\,e^{-\tau}}{(1+p)(1+\sigma_0 p)}\;
\exp\!\biggl\{
-\frac{\sigma_{\rm t}^{2}p^{2}}{\Omega_p}
\Bigl[\!\frac{w^2}{v^2}\tau
      +\Omega_p\frac{v^{2}-w^{2}}{v^{3}}\bigl(1-e^{-v\tau/\Omega_p}\bigr)\Bigr]\!
\biggr\}.
\end{align}
\end{widetext}
The variational problem defined by Eqs.~\eqref{eqn:upper_bound_int}-\eqref{I-F} generalize the all-coupling Feynman solution of polaron problem to the case of 2D polar materials. Minimization of the right–hand side with respect to the variational
parameters $v$ and $w$ yields the Feynman estimate $E_{\rm F}$ of the ground state energy of the 2D Fr\"ohlich polaron. 

It is easy to see that in the weak- and strong-coupling regimes, the Feynman solution recovers the results of Sec.~IV. 
In the limit $\alpha_{\rm m}\ll 1$, the right-hand side in Eq.~\eqref{eqn:upper_bound_int} is dominated by the first term, which is minimized when $v=w$. The polaron energy is then given by the second term evaluated at this point. By setting $v=w$, and performing the trivial integration over $\tau$ in Eq.~\eqref{I-F}, we find,
\begin{align}
    {\mathcal I}_{\rm F}(\sigma_{0},\sigma_{\rm t};v,v) =
    \mathcal{I}_0(\sigma_{0},\sigma_{\rm t}),
\end{align}
where $\mathcal{I}_0(\sigma_{0},\sigma_{\rm t})$ is defined by Eq.~\eqref{eqn:int_wc_dimless} that is the weak-coupling result derived in Sec.~IV~A.

In the opposite, strong‑coupling regime $\alpha_{\rm m}\gg 1$, the phonon cloud is tightly bound to the electron. Variationally, this corresponds to $w=0$ with $v\gg 1$. In this case, the first term in Eq.~\eqref{eqn:upper_bound_int} reduces to $\hbar\omega_{\rm t}\,v/2$, whereas the integral in Eq.~\eqref{I-F} simplifies as, 
\begin{align}
    {\mathcal I}_{\rm F}\approx
\int\limits_{0}^{\infty}\!{\rm d}p\;\frac{\sigma_{\rm t}(\sqrt{\sigma_0}+1)^2\,p\,e^{-\frac{\sigma_{\rm t}^2 p^2}{v}}}{(1+p)(1+\sigma_0 p)}.
\end{align}
Identifying $v=\hbar/(m_{e}\omega_{\rm t}a^{2})$ we observe that in this limit the upper bound of Eq.~\eqref{eqn:upper_bound_int} coincides with the Landau–Pekar functional~\eqref{eq:LP-Gauss-int} obtained in Sec.~IV~B for the Gaussian trial function.

Apparently, for intermediate couplings $0<\alpha_{\rm m}<\infty$, Eqs.~\eqref{eqn:upper_bound_int}-\eqref{I-F} provide a smooth interpolation between the above two limits. 
In general, the effective 2D coupling constant $\alpha_{\rm m}$ defined in Sec.~III~D, enters the energy of Eq.~\eqref{eqn:upper_bound_int} only as a prefactor in front of a dimensionless integral ${\mathcal I}_{\rm F}(u,v)$ that depends on the variational parameters $v$ and $w$. However, in contrast to the 3D case, the integral ${\mathcal I}_{\rm F}$ defined in Eq.~\eqref{I-F} also contains the remaining two material parameters $\sigma_{0}$ and $\sigma_{\rm t}$, which is a specific 2D feature.

\begin{figure}[t]
    \centering
    \includegraphics[width=\linewidth]{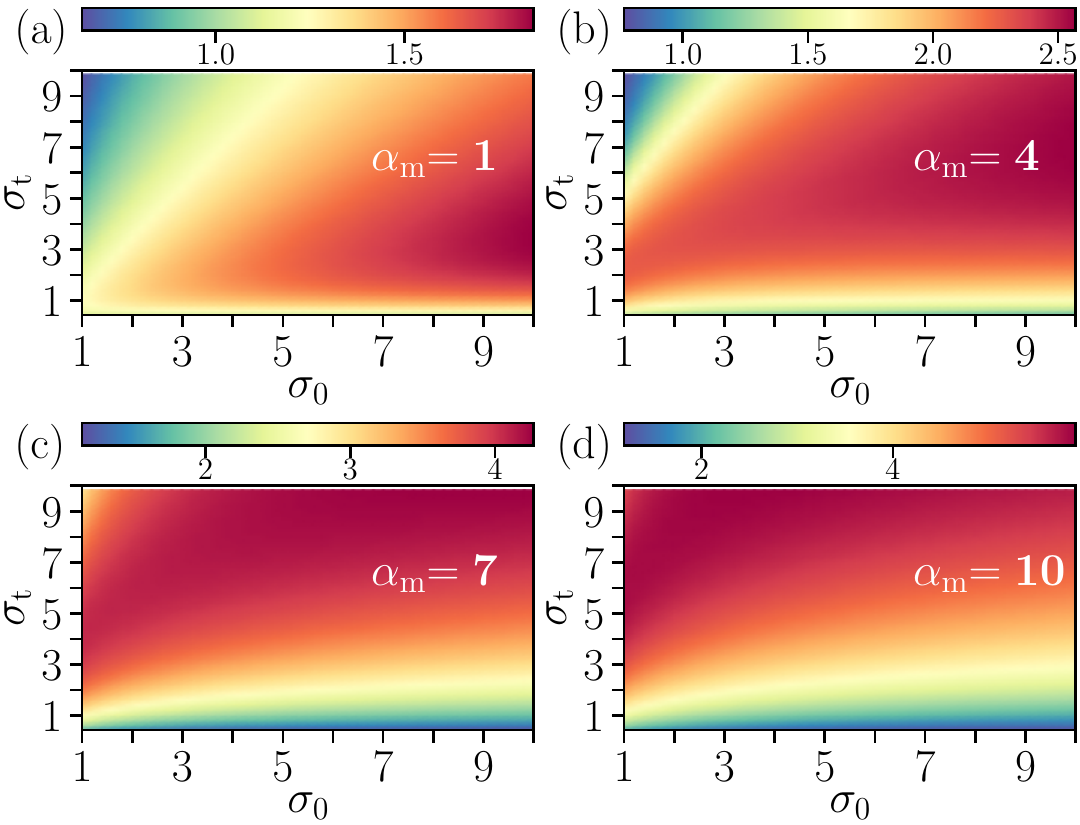}
    \caption{%
        Normalized variational energy $E^e_{\rm F}/E^e_{\rm F}(\sigma_0=1,\sigma_{\rm t}=0.5)$
        for four values of the coupling prefactor~$\alpha_{\rm m}$:
        (a)~$\alpha_{\rm m}=1$, 
        (b)~$\alpha_{\rm m}=4$,
        (c)~$\alpha_{\rm m}=7$, and 
        (d)~$\alpha_{\rm m}=10$.
        The color scale (different in each panel) indicates the
        magnitude of ${\mathcal I}_{\rm F}$ over the plane spanned by the
        static‑to‑high‑frequency dielectric‑length ratio
        $\sigma_{0}$
        and the transverse‐phonon length
        $\sigma_{\rm t}$.
        For weak coupling the integral remains close to unity,
        reproducing the perturbative behavior,
        whereas stronger coupling progressively amplifies its
        dependence on the material parameters.}
    \label{fig:heat_map_4panels}
\end{figure}
In Sec.~III~A, we have demonstrated that the strongest dependence on the material-specific parameters is encoded in the effective coupling constant, while the variation of polaron energy over the $(\sigma_0,\sigma_{\rm t})$-plane is moderate. This tendency is preserved for larger coupling.  In Fig.~\ref{fig:heat_map_4panels}, we illustrate the dependence of the variational polaron energy obtained from Eq.~\eqref{eqn:upper_bound_int} depends on $\sigma_0$ and $\sigma_{\rm t}$. Each panel displays a color map of $E_{\rm F}/E_{\rm F}(\sigma_0=1,\sigma_{\rm t}=0.5)$ for four representative coupling strengths $\alpha_{\rm m}=1,\,4,\,7,$ and $10$. We observe that all panels look very similar, with a slight tendency of increasing the variations over the $(\sigma_{0},\sigma_{\rm t})$-plane at larger couplings. In all panels in Fig.~\ref{fig:heat_map_4panels} over the whole shown region, the polaron energy changes at most by a factor of 2. On the other hand, it increases by more than an order of magnitude when $\alpha_{\rm m}$ goes from 1 to 10 (see next section).  Therefore, even well beyond the weak coupling regime, the effective Fr\"ohlich constant $\alpha_{\rm m}$  along provides a good overall characterization of the coupling strength.  However, for the quantitative description of the 2D polaron the dependence on the other parameters becomes important, especially when the strong coupling physics comes into play.

%------------------------------------------------------------
\subsection{Renormalization of the polaron mass}
\label{sec:mass-renorm}
%------------------------------------------------------------

In order to obtain the dynamical mass of the polaron, we follow the method introduced in~\cite{feynman1955slow}. We set the whole electron–phonon complex in motion at a small constant velocity $\mathbf U$, and determine the quadratic increase in energy with $U$.  A rigid drift is implemented by shifting the electron path according to $\mathbf r_{e}(u)\!\to\!\mathbf r_{e}(u)+\mathbf U\,u$.  Under this transformation, the Euclidean action $S_{\mathrm{eff}}$ is modified in three ways: (i) the electron kinetic term acquires the classical shift $m_{e}U^{2}/2$;  (ii) every vertex factor gets an extra phase $\exp[i\mathbf k\!\cdot\!\mathbf U (u-u')]$;  (iii) the same phase appears in the trial action $S_{0}$, so that all Gaussian averages can still be performed analytically.

With the drift included, the form factor introduced above in Eq.~\eqref{form-factor}  
\begin{align}
I_{\mathbf U}(\mathbf k;u,u')=
I(\mathbf k;u,u')\,
\exp\!\bigl[i\mathbf k\!\cdot\!\mathbf U(u-u')\bigr],
\end{align}
where $I(\mathbf k;u,v)$ is given by Eq.\,\eqref{eq:I-kuv}.  Expanding the additional phase to the second order in $\mathbf U$ and noticing that the linear term vanishes after angular integration over $\mathbf k$, one finds that only the $U^{2}$ contribution appears in the variational bound.  Re‑evaluating $\langle S_{\mathrm{eff}}-S_{0}\rangle_{0}$ with the modified $I_{\mathbf U}$ yields three pieces: the $C$–term stemming from the trial action, the ground–state energy $E_{0}$ of the trial model, and the genuine electron–phonon part that contains the explicit coupling $|V_{k}|^{2}$.

The contributions that originate solely from the trial action are identical to those in Ref.~\cite{feynman1955slow} , 
\begin{multline}
\frac{C}{2}
\int\limits_{0}^{\beta\hbar}\!\!\int\limits_{0}^{\beta\hbar} 
{\rm d}u\,{\rm d}u'\,
e^{-W|u-u'|}
\Bigl(-\nabla_{\mathbf{k}}^{2} I_{\mathbf U}(\mathbf k;u,u')\Bigr)\bigl|_{\mathbf k=0}\\
=\frac{2C\beta\hbar^{2}}{m_e W V}
+\frac{C\,\mathbf U^{2}}{2}\int\limits_{0}^{\beta\hbar}\!\!\int\limits_{0}^{\beta\hbar} {\rm d}u\,{\rm d}u'\,(u-u')^{2}e^{-W|u-u'|}\\
=\frac{2C\beta\hbar^{2}}{m_e W V}
+\frac{2C\beta\hbar\,\mathbf U^{2}}{W^{3}},\qquad (\beta\to\infty),
\end{multline}
so that $\partial E_{0}/\partial C = 2\hbar/(m_{e}WV)+2U^{2}/W^{3}$ and, after integrating with the condition $E_{0}(C{=}0)=\tfrac{1}{2}m_{e}U^{2}$, one obtains  
\begin{align}
E_{0}= \hbar(V-W)+\frac{U^{2}}{2}\Bigl(m_{e}+\frac{4C}{W^{3}}\Bigr).
\end{align}
The electron–phonon part can be written as  
\begin{multline}
{\mathcal F}_U=
\beta
\sum_{\mathbf k}|V_{k}|^{2}
\int\limits_{0}^{\infty}\!{\rm d}u\,
e^{-\omega_{ {\rm l},k}u}\\\times
I(\mathbf k;u,0)\left(1 -\dfrac{(\mathbf{k}\cdot \mathbf{U})^2\,u^2}{2}\right)\\={\mathcal F}-\beta\hbar \dfrac{\delta m U^2}{2},
\end{multline}
where we expanded $\exp[\,i\,\mathbf{k} \cdot \mathbf{U}\,(t -s)]$ and taking into account carrying out the angular average over $\mathbf k$ kept only two first non-trivial terms. Collecting all contributions, we obtain,
\begin{align}
    E(U)\leq \dfrac{m_e^*\mathbf{U}^2}{2}+\hbar\omega_{\rm t}\dfrac{(v-w)^2}{2v}-\dfrac{1}{\beta\hbar}{\mathcal F}.
\end{align}
The electron mass, renormalized by polaron effects, can be read out of the first term,
\begin{align}
    m_e^*=m_e+\delta m
\end{align}
where the mass correction is
\begin{align}
\delta m(v,w)=\alpha_{\rm m}\,  m_e \, \mathcal{M}_{\rm F}(\sigma_{0},\sigma_{\rm t};v,w)
\end{align}
with
\begin{widetext}
\begin{align} \label{M-F}
\mathcal{M}_{\rm F}(\sigma_{0},\sigma_{\rm t};v,w)=
\int\limits_{0}^{\infty}\!
{\rm d}\tau
\int\limits_{0}^{\infty}\!{\rm d}p\;\frac{\sigma_{\rm t}^3(\sqrt{\sigma_0}+1)^2\,p^3\tau^2\,e^{-\tau}}{(1+\sigma_0 p)^2}\;
\exp\!\biggl\{
-\frac{\sigma_{\rm t}^{2}p^{2}}{\Omega_p}
\Bigl[\!\frac{w^2}{v^2}\tau
      +\Omega_p\frac{v^{2}-w^{2}}{v^{3}}\bigl(1-e^{-v\tau/\Omega_p}\bigr)\Bigr]\!
\biggr\}.
\end{align}
\end{widetext}
Evaluation of $\delta m$ at the optimal parameters $(v^{*},w^{*})$  minimizing the upper bound of Eq.~\eqref{eqn:upper_bound_int}, finally yields the polaron effective mass.

To conclude this section, we note that in the weak coupling regime $v^*=w^*$, and the integral over $\tau$ in Eq.~\eqref{M-F} can be easily calculated. By performing the integration, we recover the mass renormalization obtained recently using the variational approach of LLP \cite{shahnazaryan2025polarons}. That is, $\mathcal{M}_{\rm F}(\sigma_0,\sigma_{\rm t};v^*,v^*)=\mathcal{M}_{\rm LLP}(\sigma_0,\sigma_{\rm t})$, where
\begin{align}
    \mathcal{M}_{\rm LLP}(\sigma_0,\sigma_{\rm t})= \int\limits_{0}^{\infty}\frac{\sigma_{\rm t}^3(\sqrt{\sigma_0}+1)^2p^3\, {\rm d}p}{(1+\sigma_0 p)^2(1+\sigma_{\rm t}^2p^2/\Omega_p)^3} .
\end{align}

\section{Numerical results}
\label{sec:results}
\begin{table*}[t]
\caption{Polaron binding energies and effective masses for selected monolayers.
For each carrier (electron or hole) we list the binding energy obtained
in the weak‑coupling approximation ($E_{0}$), the
strong‑coupling/Landau–Pekar approximation with a \emph{Gaussian} trial
state ($E_{\rm LP}$), and the Feynman variational result ($E_{\rm F}$),
all in meV. Material parameters taken from experimental and first-principal calculations data for a set of 2D polar materials: TO phonon energy $\hbar \omega_{\rm t}$, 
screening lengths $r_0$, $r_\infty$\footnote{the screening lengths are extracted from Ref. \cite{sio2023polarons} as
$r_{0[\infty]} = \epsilon_{0[\infty]} d/2$, 
where $\epsilon_{0[\infty]}$ is the static [high-frequency] dielectric constant, $d$ is the monolayer thickness.}, 
electron and hole effective masses $m_e$, $m_h$ (taken from Ref. \cite{sio2023polarons}).
} 
\label{table:1}
\begin{ruledtabular}
\begin{tabular}{lcclccccc}
Mat. & $E_{\rm LP}^e/E_0^e/E^{e}_{\rm F}$ (meV)& $E_{\rm LP}^h/E_{0}^h/E_{\rm F}^h$ (meV)& $m_e/m_e^{\rm LLP}/m_e^{\rm F}$&$m_h/m_h^{\rm LLP}/m_h^{\rm F}$&$\alpha_{\rm m}^e/\alpha_{\rm m}^h$ & $\hbar\omega_{\rm t}$ (meV)  & $r_0$ (nm)    &$r_\infty$ (nm)       \\[3pt]\hline
hBN & $15.00/129.3/131.6$   & $9.433/116.3/118.2$  &$0.83$$/1.00/1.04$ &$0.65$$/0.77/0.80$ & $0.65/0.56$ & $172.3$\cite{sohier2017breakdown}            &$1.076$ \cite{sohier2017breakdown}        &$0.780$ \cite{sohier2017breakdown}        \\
GaN  & $0.647/55.16/56.4$ &$44.78/132.0/138.4$   &$0.24$$/0.29/0.30$ &$1.35$$/2.02/2.49$ &$0.64/1.51$& $73.30$\cite{Sanders2017}  &$1.109$        &$0.756$    \\
AlN  & $12.52/105.4/110.1$ & $80.98/186.5/201.4$   &$0.51$$/0.71/0.83$ &$1.49$$/2.53/3.99$ & $1.19/2.03$ & $74.15$\cite{doi:10.1021/acs.jpcc.6b09706}   &$0.763$        &$0.466$  \\
HfSe$_2$ & $71.55/89.56/101.6$ & $81.82/98.51/112.2$  &$0.18$$/0.43/1.26$ &$0.23$$/0.57/1.78$ & $5.75/6.50$ &$11.10$\cite{Li_2024} &$21.75$ &$4.246$   \\
HfS$_2$ & $92.04/123.2/137.7$ & $127.9/155.1/174.6$  &$0.24$$/0.54/1.32$ &$0.44$$/1.07/3.02$ & $5.00/6.77$& $17.81$\cite{doi:10.1021/acsomega.1c04286}   &$13.98$        &$2.974$  \\
ZrS$_2$ & $89.92/142.9/153.2$& $80.58/133.7/143.2$   &$0.31$$/0.58/0.93$ &$0.26$$/0.48/0.74$  & $3.80/3.48$  &$29.82$\cite{Pandit_2021} &$10.62$ &$2.826$  \\
\end{tabular}
\end{ruledtabular}
\end{table*}

\begin{figure}
    \centering
    \includegraphics[width=1\linewidth]{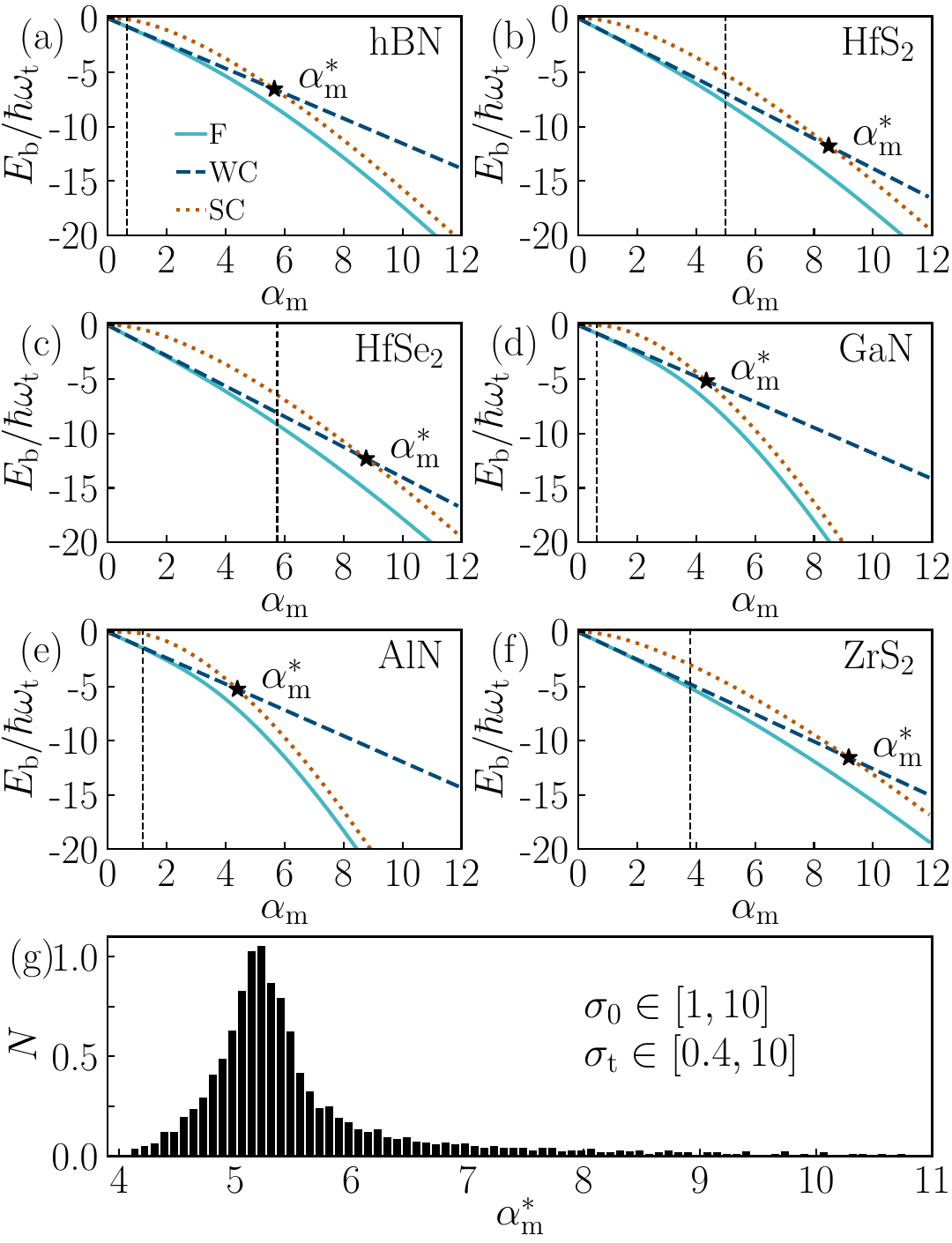}
    \caption{%
        (a)-(f): binding energy as a function of the coupling constant
        $\alpha_{\rm m}$ for six pairs of parameters $(\sigma_0,\sigma_{\rm t})$ corresponding to indicated polar monolayers.
        Solid turquoise: Feynman variational result;
        dashed blue: weak‑coupling approximation;
        dotted orange: strong‑coupling approximation with the Gaussian trial function.
        The vertical dashed line indicates the  physical value
        $\alpha_{\rm m}^e$ for each material. The star marks the crossover
        $\alpha_{\rm m}^{\ast}$ where the strong‑coupling branch overtakes
        the weak‑coupling one.
        (g): histogram of $\alpha_{\rm m}^{\ast}$ obtained from a
        random scan over the region 
        $\sigma_{0}\!\in\![1,10]$ and
        $\sigma_{\rm t}\!\in\![0.4,10]$ in the $(\sigma_0,\sigma_{\rm t})$-plane.}
    \label{fig:dif_mat_pan}
\end{figure}

The variational formalism developed in Sec.~V can be applied to real materials, provided that the parameters characterizing a polar monolayer -- the energy $\hbar\omega_{\rm t}$ of TO phonons, the static and high‑frequency screening lengths $r_{0}$ and $r_{\infty}$, and the electron (or hole) mass $m_{e(h)}$ -- are specified. In Table~\ref{table:1}, we summarize the input data collected from the literature for a representative set of 2D polar materials, together with the corresponding results for the energies and effective masses of polarons.  
For each material, we show the binding energy of the electron and hole
polarons in three approximations: the weak‑coupling limit
($E^{e(h)}_{0}$) which gives results identical to LLP approximation \cite{shahnazaryan2025polarons}, the strong‑coupling/Landau–Pekar limit
($E^{e(h)}_{\rm LP}$), and the full Feynman variational result
($E^{e(h)}_{\rm F}$).  The same hierarchy is reported for the effective masses:
the band mass $m_{e(h)}$, the polaron mass calculated within the weak‑coupling/LLP approach
$m_{e(h)}^{\rm LLP}$, and the variational mass
$m_{e(h)}^{\rm F}$. All strong coupling results in Table~\ref{table:1} are obtained with the Gaussian trial wave function because, compared to the hydrogenic ansatz, it systematically produces lower variational energies (stronger binding). The ratio
$E_{\rm LP}^{\rm G}/E_{\rm LP}^{\rm H}$ equals 
$1.248$ for hBN, $1.121$ for HfS$_2$, $1.117$ for HfSe$_2$, $1.408$ for GaN, $1.304$ for AlN, and $1.123$ for ZrS$_2$.

The results presented in Table~\ref{table:1} show that for materials with a small effective coupling constant, such as hBN and GaN in which $\alpha_{\rm m}\!\lesssim\!1$, the Feynman energies are within a few percent of the weak coupling predictions, while the Landau–Pekar approximation dramatically, up to an order of magnitude, underestimates the binding energy.  Materials with a larger difference between the static and high-frequency screening lengths, and softer phonons, such as, for example, 
HfSe\textsubscript{2}, HfS\textsubscript{2}, and ZrS\textsubscript{2}, have a larger effective coupling constant $\alpha_{\rm m}\sim 4-5$. In these materials, the accuracy of the weak coupling results worsens, being within $5$–$10$\,\% of the Feynman estimate. On the other hand, the quality of the Landau-Pekar energy improves, but for all studied materials the weak-coupling/LLP gives better results than the adiabatic strong-coupling theory. 
In all cases, including weakly coupled materials, the weak-coupling/LLP results for the polaron mass are less accurate than those for the binding energy. In fact, the deviation of the LLP and Feynman estimates for the effective mass ranges from $4$\,\% in hBN to more than a factor of three in HfS\textsubscript{2}.

To illustrate the dependence of the polaron binding energy on the effective coupling constant $\alpha_{\rm m}$, we take six pairs of parameters $(\sigma_0, \sigma_{\rm t})$ corresponding to specific materials from Table~\ref{table:1}. Then, for each pair, we plot the polaron energy, obtained from the accurate all-coupling variational theory of Sec.~V, as a function of $\alpha_{\rm m}$, together with the results of the weak and strong coupling approximations. Figure~\ref{fig:dif_mat_pan} summarizes these results for the case of electron polaron.
Panels (a)–(f) correspond to the pairs $(\sigma_{0},\sigma_{\rm t})$ of screening parameters for the materials indicated on the panel. The vertical dashed line shows the physical $\alpha_{\rm m}$ for each material.

The solid turquoise line is the full result of the Feynman-type variational theory, developed in Sec.~V. The dashed blue
line is the weak coupling approximation of Sec.~III~A, which can also be restated in a variational form \cite{shahnazaryan2025polarons}. Finally, the dotted orange line shows the
strong‑coupling Landau-Pekar energy obtained with the Gaussian trial wave function. The asterisk indicates the intersection of the weak and strong coupling curves, while $\alpha_{\rm m}^{\ast}$ is the crossover value of the coupling constant above which the strong coupling ansatz variationally wins over the weak coupling approximation and becomes closer to apparently the best result of the Feynman theory.
The lower panel shows the distribution of the crossover coupling constant $\alpha_{\rm m}^{\ast}$ obtained by scanning over the broad ranges of screening parameters $\sigma_{0}\!\in\![1,10]$ and $\sigma_{\rm t}\!\in\![0.4,10]$ in the $(\sigma_{0},\sigma_{\rm t})$-plane.
The distribution clearly shows that most of the systems cluster around $\alpha_{\rm m}^{\ast}\!\approx\!5$. Remarkably, this value is about twice the smaller than the crossover value of $\alpha^\ast_{\rm 3D}=3\pi\approx 10$ for bulk 3D polarons, but somewhat larger than the threshold coupling constant $\alpha^\ast_{\rm surf}=4$ for surface polarons. Therefore, in a sense, the Fr\"ohlich polarons in genuine 2D materials take an intermediate position between the bulk and surface polarons in 3D polar crystals.

%%%%%%%%%%%%%%%%%%%%%%%%%%%%%%%%%%%%%%%%%%%%%%%%%%%%%%%%%%%%%%%%%%%%%%
\section{Conclusion}
We presented a macroscopic
theory of Fr\"{o}hlich polarons in strictly 2D
polar crystals. By construction, in our approach the quantitative description of polaron physics does not involve any adjustable parameters but requires only the knowledge of experimentally accessible macroscopic polarizabilities and effective masses. The required parameters can also be routinely obtained using standard first-principle methods. Therefore, our macroscopic theory can be naturally incorporated into a fully {\it ab initio} multiscale computational scheme.

Starting from a long wavelength continuum description of optical phonons in 2D, we derived an effective
Lagrangian LO phonons, interacting with electrons. It turns out that in 2D polar materials the electron-LO phonon interaction is characterized by three independent dimensionless parameters. We have identified one of them, $\alpha_{\rm m}$ in Eq.~\eqref{alpha-m}, as an effective coupling constant that provides a natural measure of interaction strength and can be viewed as a 2D generalization of the standard Fr\"ohlich constant. The dependence on the remaining two parameters appears to be moderate, whereas it becomes more important quantitatively at stronger coupling. 

Aimed at a high-quality description at arbitrary coupling strength, we generalized the Feynman path integral variational theory to our 2D system, derived the two-parameter variational functional for the polaron energy, and the generalization of Feynman result for the polaron mass. By analyzing the dependence of the polaron energy on the effective coupling strength, we identified the value of $\alpha_{\rm m}^{\ast}\!\approx\!5$ above which the weak coupling regime crosses over into the regime of strongly coupled polarons. However, none of the 2D materials currently known and typically discussed in the literature \cite{sio2019ab, shahnazaryan2025polarons} has the coupling constant that exceeds the strong-to-weak coupling threshold. In most cases, the weak coupling approach \cite{shahnazaryan2025polarons} provides reasonable estimates of the polaron energies, close to the Feynman variational results. However, the polaron mass shows a much stronger dependence on the coupling strength. In fact, for some materials studied in this work, we observed that the weak coupling approximation gives a polaron mass that is up to three-four times smaller compared to the accurate variational estimate. 

In conclusion, the present framework provides a transparent and accurate tool for analyzing charge transport, optical response, and many‑body phenomena in polar monolayers.
Extensions to finite temperature, multiple optical branches, and more complex quasiparticles, such as exciton‑polarons or trions dressed by LO phonons, can be developed along similar lines and will be the subject of future work.

\textit{Acknowledgments}.
The work of A.K. is supported by the Icelandic Research Fund (Ranns\'oknasj\'o{\dh}ur, Grant No.~2410550). 
V.S. acknowledges the support of “Basis” Foundation (Project No. 25-1-3-11-1). 
IVT acknowledges support from the Spanish MCIN/AEI/
10.13039/501100011033 through the project PID2023-148225NB-C32, and the Basque Government (Grant No. IT1453-22).

\bibliography{refs}

\end{document}